# On The Performance of a Two-Sided Shewhart Chart for Continuous Proportions with Estimated Parameters

Athanasios C. Rakitzis[1]

Department of Statistics and Insurance Science, University of Piraeus, Piraeus, Greece

**Abstract:** During the recent years there was an increased interest in studying the performance of different types of control charts, under various distributional models for continuous proportions, such as percentages and rates. In this work we consider the Kumaraswamy distribution, a popular and flexible distributional model for data in the unit interval (0,1) and investigate further the properties of a two-sided chart for individual observations for monitoring these types of processes, when the process parameters are unknown. Specifically, using Monte Carlo simulation, we evaluate the performance of the chart under a conditional perspective and provide empirical rules on how to select the appropriate size for the Phase I sample. In addition, we explore possible adjustments on the control limits of the chart, which take into account the available Phase I sample. The performance of the chart is also investigated for several out-of-control situations. The results show that for small and moderate size Phase I samples, practitioners have to choose whether they prefer a guaranteed in-control performance or an improved out-of-control performance. The implementation of the considered methods in practice is discussed via two numerical examples.

## 1. Introduction

Control charts are graphical statistical methods that are used for the statistical monitoring of a process. They are the most useful method of statistical process monitoring, originally proposed by W. A. Shewhart (1926). The Shewhart-type control chart is the simplest and most often used control chart and applies for monitoring a variety of processes, either industrial or non-industrial (see, for example, Aykroyd et al. (2019), Tran et al. (2022), Pérez-Benítez et al. (2023), Arciszewski (2023) and references therein).

---
[1] Corresponding author. Email: arakitz@unipi.gr

To set up a control chart, samples of size $n \geq 1$ are obtained at each sampling stage $t$, $t = 1, 2, ...$ and the value of an appropriate statistic is calculated and plotted on the chart. A critical-to-quality characteristic (CTQ) $X$ is related to each process and values from the distribution of $X$ are obtained and form the sample. The common assumption is that $X$ follows a normal distribution with mean $\mu$ and variance $\sigma^2$, i.e., $X \sim \mathcal{N}(\mu, \sigma^2)$. The most often used control charts such as the $\bar{X}$ and $S$ charts or the $I$- and $MR$-charts, are constructed and applied under the assumption of normality. See, for example, Chapter 6 in Montgomery (2020). However, there are several cases where this assumption is violated and consequently, control charts based on the appropriate distributional model for $X$ need to be established and applied.

When the support of the distribution of $X$ is the unit interval $(0,1)$, there are several candidate models. The most common model is the Beta distribution (see, for example, Kieschnick and McCullough (2003), Gupta and Nadarajah (2004)), which is quite popular due to its flexibility in modeling data with different characteristics. A closely related model to that of Beta distribution is the Kumaraswamy distribution (Kumaraswamy (1980)), with several applications in hydrological and environmental data (see, for example Bayer et al. (2017), Sagrilo et al. (2021)). Therefore, the Kumaraswamy distribution can be used for modelling continuous proportions, i.e., values in $(0,1)$ such as percentages and proportions which are not results from Bernoulli experiments; for example, the proportion of crude oil converted to gasoline after distillation and fractionation (Prater 1956), the daily relative humidity in a city (Raschke 2011), the ratio of premiums plus uninsured losses divided by total assets of a company (Mazucheli et al. 2020) or the percentage of a specific ingredient in a pharmaceutical product (Espinheira and Silva, 2020).

Control charts based on the Kumaraswamy distribution have been studied by Lima-Filho et al. (2020) and Lima-Filho and Bayer (2021). In both works, the focus is on Shewhart charts with probability limits. However, the authors did not consider the estimation effect on the design and performance of the considered charts, even though Lima-Filho and Bayer (2021) considered the case of estimated control limits. However, they did not investigate how the different estimates affect the performance of the chart, whereas they did not provide any guidance regarding the size of the Phase I sample or how to determine the control limits of the chart in order to have a guaranteed in-control performance. In addition, both Lima-Filho et al. (2020) and Lima-Filho and Bayer (2021) considered only the case of shifts in the in-control process mean level.

Motivating by this research gap, in this work we investigate the differences in the performance of a two-sided Shewhart-type chart for continuous proportions between the case of known process parameters (or Case K) and the case of unknown process parameters are unknown (or Case U). The aim is to suggest ways to mitigate this difference, either by proposing the size of the Phase I sample or by providing adjustments on the control limits of the chart, given that the size of the Phase I sample is pre-specified.

The literature of control charts with estimated parameters is quite extended with several suggested approaches and findings. However, the majority of the related works focus on the case of normal distribution for $X$ (see, for example, Atalay et al. (2020), Does et al. (2020), Jardim et al. (2020), Mosquera et al. (2021), Sarmiento et al. (2022) and references therein) while for recent works for non-normal data, we refer to Kumar (2022), Madrid-Alvarez et al. (2024), Huang (2022) and Saghir et al. (2021). However, the case of control charts with estimated parameters for continuous proportions has not been explored yet, to the best of our knowledge.

The structure of this paper is as follows. In Section 2, we provide in brief the properties of the Kumaraswamy distribution and the related Shewhart-type chart with probability limits, in Case K and Case U. Also, we discuss the most common performance measures for assessing the performance of the chart in Case U and compare it with the performance in Case K. In Section 3 we present the results of a simulation study regarding the performance of the considered chart. In addition, we provide adjustments on the control limits of the chart so as the chart has the desired IC performance in Case U. Furthermore, we study the effect of these adjustments in the OOC performance of the chart when a shift in exactly one of the process parameters occurs. The practical implementation of the proposed methods is demonstrated with two examples in Section 4 while in Section 5, we summarize the finding of this research and suggest topics for future research.

## 2 Materials and Methods

In this section we present, in brief, the basic properties of the Kumaraswamy distribution and then we proceed with the two-sided Shewhart chart, namely the $SH_K$-chart, in the case of known parameters (or Case K) as well as in the case of unknown parameters (or Case U).

## 2.1 The Kumaraswamy Distribution

Let $Y$ be a random variable (r.v.) which follows a Kumaraswamy distribution with parameters $\theta_1, \theta_2$ (i.e., $Y \sim Kuma(\theta_1, \theta_2)$). Then, its probability density function (pdf) is given by (see e.g., Jones (2009)):

$$f_K(y; \theta_1, \theta_2) = \theta_1 \theta_2 y^{\theta_1 - 1} (1 - y^{\theta_1})^{\theta_2 - 1}, \quad 0 < y < 1 \tag{1}$$

where $\theta_1, \theta_2 > 0$ are the shape parameters of the distribution. The expected value and the variance of $Y$ are given by:

$$E(Y) = \theta_2 B\left(1 + \frac{1}{\theta_1}, \theta_2\right), \quad V(Y) = \theta_2 B\left(1 + \frac{2}{\theta_1}, \theta_2\right) - \left[\theta_2 B\left(1 + \frac{1}{\theta_1}, \theta_2\right)\right]^2 \tag{2}$$

where $B(a, b) = \int_0^1 t^{a-1}(1-t)^{b-1} dt$ is the Beta function.

Also, the cumulative distribution function (cdf) and the inverse cdf are, respectively, given by:

$$F_K(y; \theta_1, \theta_2) = P(Y \leq y) = \int_0^y f_K(t; \theta_1, \theta_2) dt = 1 - (1 - y^{\theta_1})^{\theta_2}, \quad 0 < y < 1 \tag{3}$$

$$F_K^{-1}(u; \theta_1, \theta_2) = \left[1 - (1 - u)^{\frac{1}{\theta_2}}\right]^{1/\theta_1}, \quad 0 < u < 1 \tag{4}$$

The $F_K^{-1}(u; \theta_1, \theta_2)$ is the $u$-percentile point of the $Kuma(\theta_1, \theta_2)$ distribution. It is worth noting (see Jones 2009) that the $F_K(y; \theta_1, \theta_2)$ and $F_K^{-1}(u; \theta_1, \theta_2)$ are given in closed and (relatively) simple formulas which is considered as an advantage of the Kumaraswamy distribution over other distributions defined in (0, 1) such as the Beta distribution.

## 2.1 The SH$_K$-Chart with Known Parameters

Let $a \in (0,1)$ be the probability of false alarm, i.e., the probability the chart gives an OOC signal when the process is actually IC. We want to monitor the process, whose output is modelled according to a Kumaraswamy distribution, by using a two-sided Shewhart-type chart for individual observations with equal tail probability limits. Specifically, the lower and the upper control limits $LCL, UCL$ satisfy the following equations:

$$P(Y \leq LCL) = a/2 \text{ and } P(Y > UCL) = a/2 \tag{5}$$

where $Y \sim Kuma(\theta_{01}, \theta_{02})$. Let also $\theta_{01}, \theta_{02} > 0$ be the IC values of the shape parameters $\theta_1, \theta_2$ of the process. Therefore,

$$\alpha = 1 - P(LCL \leq Y \leq UCL | \boldsymbol{\theta} = \boldsymbol{\theta}_0), \tag{6}$$

where $\boldsymbol{\theta}_0 = (\theta_{01}, \theta_{02})$. Using the percentile points of the $Kuma(\theta_{01}, \theta_{02})$ distribution, the equal tail probability limits of the chart are

$$LCL = F_K^{-1}(a/2; \theta_{01}, \theta_{02}), \quad UCL = F_K^{-1}(1 - a/2; \theta_{01}, \theta_{02}) \tag{7}$$

For the center line $CL$ of the chart we can use either the IC mean of the process, that is, $CL = \theta_{02} B\left(1 + \frac{1}{\theta_{01}}, \theta_{02}\right)$, or the IC median, i.e., $CL = F_K^{-1}(0.5; \theta_{01}, \theta_{02})$. If the distribution of $Y$ is not symmetric, we suggest the use of the median. Otherwise, we can use the IC mean of the process. Obviously, this is a Phase II control chart because the values $\theta_{01}, \theta_{02}$ are assumed known. We will refer to this chart as the SH$_K$-chart. Since the SH$_K$-chart is a two-sided Shewhart chart, it gives an out-of-control signal when a single point is above the upper control limit or below the lower control limit, i.e., it gives an OOC signal at the $t^{\text{th}}$ point, $t = 1, 2, ...$, if $Y_t \notin [LCL, UCL]$.

## 2.2 The SH$_K$-Chart with Unknown Parameters

When the values $\theta_{01}, \theta_{02}$ are unknown, prior to start the process monitoring, they must be estimated from a Phase I sample. Let $X_1, X_2, ..., X_m$ be a Phase I sample from the $Kuma(\theta_{01}, \theta_{02})$ distribution. For given values $x_1, x_2, ..., x_m$, the log-likelihood function $\ell(\theta_{01}, \theta_{02}) = \log(L(\theta_{01}, \theta_{02}|\mathbf{x}))$ is

$$\ell(\theta_{01}, \theta_{02}) = m \log(\theta_{01} \theta_{02}) + \theta_{01} \sum_{i=1}^{m} \log(x_i) + (\theta_{02} + 1) \sum_{i=1}^{m} \left(1 - x_i^{\theta_{01}}\right) \quad (8)$$

The maximum likelihood estimators (MLEs) $\hat{\theta}_1, \hat{\theta}_2$ of $\theta_{01}, \theta_{02}$ are obtained by maximizing numerically the function in equation (8). In R (R Core Team (2024)) this is done with the use of `fitdistplus` package (Delignette-Muller and Dutang (2015)). Further details on estimation methods for the Kumaraswamy distribution can be found in the works of Jones (2009), Lemonte (2011) and Wang et al. (2017).

Given the estimates of $\theta_{01}, \theta_{02}$, the (estimated) control limits are obtained by directly substituting the unknown values of the parameters. Thus,

$$\widehat{LCL} = F_K^{-1}(a/2; \hat{\theta}_1, \hat{\theta}_2), \quad \widehat{UCL} = F_K^{-1}(1 - a/2; \hat{\theta}_1, \hat{\theta}_2) \quad (9)$$

We will refer to these limits as *plug-in* limits. These limits are used instead of the theoretical limits (see equation (7)) and the rule for declaring a process as out-of-control is the same, whether the limits are estimated or not.

## 2.3 Performance Measures

Usually, the performance of a control chart is evaluated in terms of the average run length ($ARL$) which is the most common measure of performance for a control chart and is defined as the expected number of points plotted on the chart until it gives for the first time an OOC signal. The $ARL$ is the expected value of the distribution of the number of points plotted on the chart until it gives for the first time an out-of-control signal, known also as the run length distribution ($RL$).

When the process is in-control, we will denote $ARL$ as $ARL_0$ and for the $SH_K$-chart in Case K is equal to $ARL_0 = 1/\alpha$, where $\alpha$ is the FAR. When the process is out-of-control, we will denote $ARL$ as $ARL_1$ and for the $SH_K$-chart is equal to $ARL_1 = 1/(1-\beta)$ where

$$\beta = P(LCL \leq Y \leq UCL | \boldsymbol{\theta} = \boldsymbol{\theta}_1), \tag{10}$$

and $\boldsymbol{\theta}_1 = (\theta_{11}, \theta_{12})$, is the pair of the OOC values of the process parameters.

In practice the true values $\theta_{01}, \theta_{02}$ are rarely (if ever) known and therefore, they must be estimated from a preliminary Phase I sample. Let $X_1, X_2, \ldots, X_m$ be a preliminary Phase I sample where $X_i \sim Kuma(\theta_{01}, \theta_{02})$, $i = 1, 2, \ldots, m$. Using the MLE method we estimate $\theta_{01}, \theta_{02}$ and let $\hat{\theta}_1, \hat{\theta}_2$ be the respective estimates. Then we use them in equation (10) to calculate the (estimated) control limits which are then used instead of the true (theoretical) control limits. However, different Phase I samples will end up with different estimates for $\theta_{01}$, $\theta_{02}$. Thus, there is a variability between the users of the chart, known as *practitioner-to-practitioner variability*, which is attributed to the use of different Phase I samples by the different users of the chart. Also, the $ARL$ is no more an appropriate performance measure and proper measures must be used.

Next, to assess the performance of the $SH_K$-chart in Case U we consider the false alarm probability (see also Zhao and Driscoll (2016))

$$\alpha_{\hat{\boldsymbol{\theta}}|\boldsymbol{\theta}_0} = 1 - F_K(\widehat{UCL}; \theta_{01}, \theta_{02}) + F_K(\widehat{LCL}; \theta_{01}, \theta_{02}), \tag{11}$$

which is the probability that the chart gives a false signal, given the estimates of the control limits $\widehat{LCL}$ and $\widehat{UCL}$. The false alarm probability $\alpha_{\hat{\boldsymbol{\theta}}|\boldsymbol{\theta}_0}$ reflects the estimation error and it can be used to evaluate the performance of the chart in Case U. Also, it is used for comparing the performance of the chart under Case K and Case U. It is also noted that $\alpha_{\hat{\boldsymbol{\theta}}|\boldsymbol{\theta}_0}$ cannot be calculated in practice since $\theta_{01}, \theta_{02}$ are unknown.

Note also that since the $\alpha_{\hat{\boldsymbol{\theta}}|\boldsymbol{\theta}_0}$ is evaluated for given $\widehat{LCL}$ and $\widehat{UCL}$ values, then the run-length distribution of the chart in this case is a geometric one and the IC $ARL$ is calculated as $(\alpha_{\hat{\boldsymbol{\theta}}|\boldsymbol{\theta}_0})^{-1}$. We will refer to this quantity as the *conditional* IC $ARL$, or $CARL_0$, which in general is a random variable since different estimates $\hat{\theta}_1, \hat{\theta}_2$ result in different control limits and thus, in different conditional IC $ARL$ values. Therefore, in the current analysis, the distribution of the $CARL_0$ is of major importance.

## 3. Numerical Results

### 3.1 Performance of the $SH_K$-chart in Case U

In the current section we will examine the estimation effect on the *ARL* performance of the $SH_K$-chart. Next, we will try to assess this effect and suggest the required size of the Phase I sample, in order to mitigate the difference between the theoretical and the true performance of the chart.

For this purpose, we consider three different Kumaraswamy distributions (see Table 1) as possible IC models. Also, in Figure 1 we provide the pdf of these three models, which have been selected due to their differences in shape. Scenarios 1 and 3 are for a non-symmetric IC distribution, with a right tail and a left tail, respectively, while the IC distribution is Scenario 2 is almost a symmetric one.

**Table 1:** The different scenarios for the three candidate models

| Scenario | Kumaraswamy | | | |
|---|---|---|---|---|
| | $\theta_{01}$ | $\theta_{02}$ | Mean | Variance |
| 1 | 2 | 30 | 0.159814 | 0.006718 |
| 2 | 3 | 12 | 0.383049 | 0.017950 |
| 3 | 12 | 100 | 0.652578 | 0.004333 |

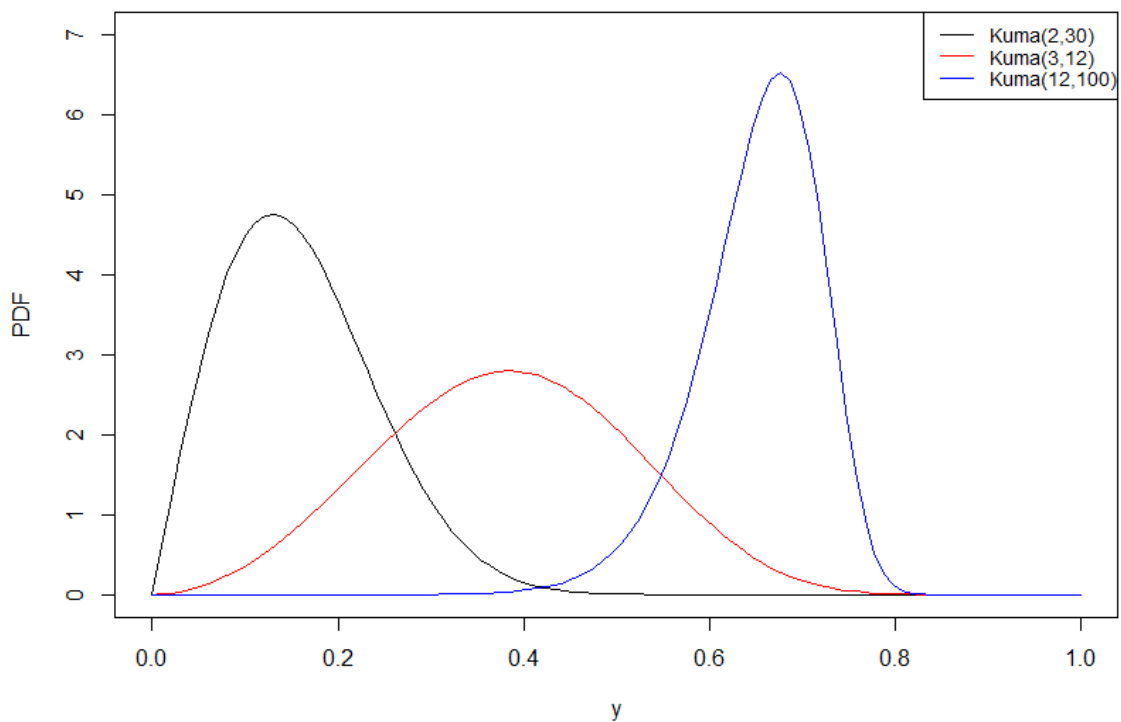

**Figure 1:** The pdf of the 3 different Kumaraswamy distributions in Table 1

Next, we provide the steps of the simulation procedure for the evaluation of the appropriate measures of performance of the chart in Case U. See also Zhao and Driscoll (2016), Mamzeridou and Rakitzis (2024) and references therein:

**Algorithm 1:** Calculation of the IC performance of the $SH_K$-chart in Case U

**Step 1:** Simulate a Phase I sample of size $m$ from the $Kuma(\theta_{01}, \theta_{02})$ distribution.

**Step 2:** Using the MLE method, estimate $\theta_{01}, \theta_{02}$ and let $\hat{\theta}_1, \hat{\theta}_2$ be the respective estimates.

**Step 3:** Calculate the plug-in control limits (using equation (9)):
$$\widehat{LCL} = F_K^{-1}(\alpha/2; \hat{\theta}_1, \hat{\theta}_2), \quad \widehat{UCL} = F_K^{-1}(1 - \alpha/2; \hat{\theta}_1, \hat{\theta}_2)$$

**Step 4:** Given the values of $\widehat{LCL}, \widehat{UCL}$ calculate the $CARL_0$ as:
$$\left(\alpha_{\hat{\theta}|\theta_0}\right)^{-1} = \left(1 - F_K(\widehat{UCL}|\theta_{01}, \theta_{02}) + F_K(\widehat{LCL}|\theta_{01}, \theta_{02})\right)^{-1},$$

**Step 5:** Repeat Steps 1-4 $N$ = 25000 times and calculate the sample mean of the 25000 $CARL_0$ values (also known as average IC $ARL$, or $AARL$), the standard deviation of the 25000 $CARL_0$ values (also known as the Standard Deviation of the IC ARL, or SDARL) as well as the empirical percentage of the cases $\{CARL_0 < ARL_0\}$, where $ARL_0 = 1/\alpha$.

In Table 2 we provide the results of the previously described simulation procedure for $\alpha = 0.0027$ (thus, the nominal, or theoretical, IC $ARL$ is 370.4) and $m \in \{50, 100, 200, 500, 800, 1000, 2000, 5000\}$. The results show the distribution of the $CARL_0$ in terms of its expected value and standard deviation ($AARL$ and $SDARL$, respectively) along with the (estimated) $\gamma$-percentile points, with $\gamma \in \{0.05, 0.10, 0.25, 0.50, 0.75, 0.90, 0.95\}$. These percentiles are useful to understand the variability in the distribution of $CARL_0$. Also, in column '*perc*' we provide the empirical (estimated) percentage of cases $\{CARL_0 < 370.4\}$, i.e. the percentage of charts with actual IC $ARL$ lower than the nominal value 370.4. For the simulation of the values from the Kumaraswamy distribution we used the package VGAM (Yee (2021)) in R and the function rkuma(...).

**Table 2:** The distribution of the $CARL_0$ when plug-in control limits are used

| $(\theta_{01}, \theta_{02})$ | $m$ | AARL | SDARL | perc | 5% | 10% | 25% | 50% | 75% | 90% | 95% |
|---|---|---|---|---|---|---|---|---|---|---|---|
| (2, 30) | 30 | 593.98 | 1412.44 | 62.95% | 38.53 | 55.66 | 107.28 | 253.16 | 536.24 | 1260.69 | 2094.11 |
| | 50 | 483.36 | 673.40 | 60.21% | 64.72 | 88.49 | 150.60 | 289.91 | 559.13 | 1038.29 | 1502.78 |
| | 100 | 421.07 | 345.24 | 57.71% | 108.76 | 136.43 | 204.98 | 325.00 | 519.10 | 795.32 | 1030.01 |
| | 200 | 392.40 | 206.22 | 55.64% | 158.05 | 189.11 | 251.02 | 350.97 | 489.32 | 665.60 | 793.61 |
| | 500 | 380.15 | 121.57 | 53.08% | 214.86 | 242.15 | 291.63 | 360.27 | 444.70 | 540.94 | 603.86 |
| | 800 | 375.65 | 93.98 | 52.92% | 244.47 | 266.27 | 308.90 | 363.81 | 429.77 | 499.99 | 548.32 |
| | 1000 | 374.92 | 83.71 | 52.44% | 256.89 | 278.34 | 315.63 | 365.70 | 422.12 | 482.66 | 522.56 |
| | 2000 | 372.47 | 58.48 | 51.96% | 285.89 | 301.85 | 331.03 | 367.39 | 408.84 | 449.46 | 475.19 |
| | 5000 | 371.24 | 36.52 | 50.94% | 314.17 | 325.78 | 346.35 | 370.11 | 394.39 | 418.99 | 434.56 |
| (3, 12) | 30 | 637.30 | 3191.46 | 62.74% | 39.56 | 56.61 | 111.35 | 247.71 | 574.13 | 1303.97 | 2075.82 |
| | 50 | 490.00 | 756.38 | 60.69% | 64.87 | 89.16 | 149.81 | 284.38 | 554.21 | 1049.37 | 1515.00 |
| | 100 | 414.92 | 331.48 | 57.61% | 110.71 | 139.63 | 206.26 | 324.79 | 514.28 | 788.21 | 1012.25 |
| | 200 | 394.93 | 204.27 | 54.76% | 161.46 | 188.82 | 252.65 | 349.04 | 483.88 | 658.04 | 782.06 |
| | 500 | 378.43 | 121.26 | 53.54% | 219.42 | 243.86 | 291.90 | 359.70 | 443.92 | 535.22 | 602.99 |
| | 800 | 376.34 | 92.84 | 52.61% | 246.35 | 268.00 | 309.70 | 364.83 | 429.71 | 498.74 | 546.68 |
| | 1000 | 374.22 | 81.73 | 52.23% | 256.94 | 277.92 | 315.70 | 365.63 | 422.47 | 483.26 | 521.72 |
| | 2000 | 373.23 | 58.39 | 51.18% | 285.89 | 301.90 | 331.84 | 368.63 | 409.35 | 450.16 | 476.06 |
| | 5000 | 371.38 | 36.47 | 51.01% | 315.09 | 325.95 | 346.21 | 369.50 | 394.47 | 418.76 | 433.70 |
| (12, 100) | 30 | 599.96 | 1829.13 | 63.14% | 39.08 | 55.97 | 109.08 | 243.33 | 575.06 | 1280.39 | 2063.97 |
| | 50 | 487.25 | 697.10 | 60.19% | 64.16 | 87.83 | 151.46 | 287.12 | 560.79 | 1049.44 | 1541.33 |
| | 100 | 419.75 | 336.30 | 57.67% | 107.89 | 136.82 | 206.76 | 325.34 | 524.35 | 802.72 | 1045.24 |
| | 200 | 394.07 | 210.20 | 55.48% | 156.55 | 185.78 | 248.56 | 345.98 | 485.04 | 658.07 | 790.54 |
| | 500 | 378.60 | 121.69 | 53.52% | 216.34 | 242.47 | 291.31 | 360.10 | 446.02 | 538.54 | 602.54 |
| | 800 | 376.14 | 94.96 | 52.34% | 241.51 | 265.33 | 308.11 | 364.99 | 430.36 | 502.60 | 549.28 |
| | 1000 | 375.08 | 84.80 | 52.45% | 254.23 | 275.05 | 314.83 | 365.38 | 424.42 | 487.26 | 528.10 |
| | 2000 | 372.39 | 59.47 | 51.98% | 283.90 | 299.92 | 330.13 | 367.68 | 409.12 | 450.83 | 478.09 |
| | 5000 | 371.51 | 37.08 | 51.11% | 313.92 | 325.68 | 346.15 | 369.29 | 395.30 | 419.82 | 434.82 |

From the results in Table 2 we notice that as $m$ increases, the $AARL$ decreases and converges to the (theoretical) value $ARL_0$ in Case K. Clearly, to reduce the impact of the estimation effect on the performance of the $SH_K$-chart, large Phase I samples, e.g. with $m > 500$, are necessary. Also, as $m$ increases the $SDARL$ decreases which means that (as expected) larger Phase I samples result in more accurate estimates and thus, the variability of the distribution of the $CARL_0$ reduces. Consequently, the differences in performance between the Case U and Case K become smaller. According to Saleh et al. (2015), practitioners can determine the size $m$ of the Phase I sample when the $SDARL$ is about 5%- 10% of the nominal IC $ARL$ value. However, the results in Table 2 show that this is achieved for $m \geq 5000$.

Furthermore, from the results in column '*perc*' we notice that for almost all the considered sizes $m$, the empirical percentages are at least 50%. Clearly, this is not expected since even for very large sample sizes almost 50% of the chart users will end up with a chart that has increased false alarm (i.e. an IC $ARL$ lower than 370.4) and thus, its IC performance will be far from the desired one. However, for $m \geq 1000$ the respective $SDARL$ values are much smaller (compared to smaller Phase I sizes). This means that as $m$ increases the distribution of the conditional IC $ARL$ tends to concentrate almost symmetrically around the nominal value $ARL_0$. Also, from the results in Table 2 we notice that for the three considered pairs $(\theta_{01}, \theta_{02})$ and for $m = 5000$, around 25% of the charts will have an IC $ARL$ lower than (approximately) 346, which differs about 7% from the nominal value 370.4.

Obviously, for smaller Phase I sample such as $m = 100, 200$ or even 500, an appropriate adjustment on the control limits is necessary, in order to achieve the desired IC performance with the available Phase I sample (i.e. when the size $m$ is pre-specified) and/or to guarantee the IC performance of the $SH_K$-chart in Case U by reducing significantly the percentage of cases $\{CARL_0 < ARL_0\}$. This is investigated in the next section.

### 3.2 The $SH_K$-chart in Case U with Adjusted Control Limits

In this section we consider two different methods for adjusting the control limits of the $SH_K$-chart in the case of estimated parameters, when the size $m$ of the Phase I sample is pre-specified. First, since the $SH_K$-chart uses equal tail probability limits, we suggest modifying the FAR and instead of using the nominal FAR $\alpha$, we suggest using $\alpha' \neq \alpha$ so as the $\frac{|AARL-ARL_0|}{ARL_0} < p$, where $ARL_0$ is calculated in Case K and equals $ARL_0 = 1/\alpha$ (nominal IC $ARL$). This means that the difference in percentage between the average of the IC $ARL$ for all charts and the nominal $ARL_0$ value does not exceed 5%. This is a very common adjustment method, which is based on the unconditional perspective (see, for example, Faria Sobue et al. (2020), Jardim et al. (2020), Kumar et al. (2020) and references therein). We will refer to this method as adjustment A (or adj-A) For the determination of $\alpha'$ we apply Algorithm 1 for a wide range of candidate $\alpha'$ values.

In Table 3 we provide the adjusted $\alpha'$ values (see the respective column) along with the $AARL$, the $SDARL$, the empirical percentage of charts with $CARL_0 < 370.4$ (column '*perc*') and percentiles of the distribution of $CARL_0$. We consider eight different sizes $m \in \{50, 100, 150, 200, 250, 300, 400, 500\}$. It is worth mentioning that this adjustment method applies for any $m$ but we did not consider sizes such as 1000 or larger, because for such sizes the variability of the distribution of the $CARL_0$ is not very large.

The results show that an increase in FAR is needed (i.e., $\alpha' > \alpha$) in order to achieve, on average, an IC performance close to the desired one. It should be mentioned that using, for example, $\alpha' = 0.0033$ instead of $a = 0.0027$ (when $m = 50$) the percentage of charts with $CARL_0 < 370.4$ exceeds 65% while even for $m = 500$, this percentage is not below 50%. However, this is expected since the adjustment A does not take into account this percentage.

**Table 3:** Adjusted FAR values (method adj-A)

| $(\theta_{01}, \theta_{02})$ | $m$ | $\alpha'$ | AARL | SDARL | perc | 5% | 10% | 25% | 50% | 75% | 90% | 95% |
|---|---|---|---|---|---|---|---|---|---|---|---|---|
| **(2, 30)** | 30 | 0.00385 | 380.73 | 753.81 | 73.79% | 32.53 | 45.68 | 84.36 | 173.70 | 387.44 | 824.66 | 1302.12 |
| | 50 | 0.00328 | 388.74 | 508.73 | 68.18% | 56.38 | 77.26 | 129.99 | 236.09 | 449.43 | 835.72 | 1203.43 |
| | 100 | 0.00291 | 388.49 | 306.20 | 61.37% | 104.77 | 131.19 | 194.48 | 305.34 | 482.04 | 733.51 | 949.09 |
| | 150 | 0.00279 | 388.55 | 246.27 | 58.82% | 132.92 | 160.64 | 225.32 | 326.40 | 482.40 | 686.88 | 836.45 |
| | 200 | 0.00274 | 386.91 | 202.41 | 56.76% | 155.86 | 184.70 | 246.86 | 341.19 | 476.18 | 643.08 | 771.40 |
| | 250 | 0.00271 | 387.83 | 178.74 | 54.79% | 172.89 | 201.36 | 263.28 | 352.07 | 472.31 | 616.27 | 725.72 |
| | 300 | 0.00267 | 388.82 | 164.20 | 53.72% | 187.89 | 215.47 | 273.94 | 357.45 | 468.51 | 601.92 | 698.09 |
| | 400 | 0.00267 | 387.63 | 141.20 | 51.85% | 205.20 | 232.47 | 288.19 | 364.10 | 460.41 | 569.36 | 651.53 |
| | 500 | 0.00264 | 388.82 | 125.03 | 50.08% | 221.94 | 248.52 | 298.26 | 370.20 | 458.12 | 552.53 | 622.96 |
| **(3, 12)** | 30 | 0.00385 | 382.31 | 792.02 | 73.88% | 32.55 | 45.39 | 83.54 | 174.37 | 384.82 | 825.57 | 1329.06 |
| | 50 | 0.00328 | 388.45 | 503.61 | 67.54% | 59.28 | 78.63 | 132.12 | 242.05 | 453.19 | 813.81 | 1194.82 |
| | 100 | 0.00291 | 388.28 | 315.30 | 61.96% | 101.95 | 128.62 | 192.05 | 299.69 | 477.98 | 744.58 | 955.47 |
| | 150 | 0.00278 | 387.91 | 240.31 | 58.56% | 136.69 | 165.45 | 226.84 | 328.60 | 479.80 | 679.09 | 839.20 |
| | 200 | 0.00272 | 388.63 | 204.47 | 56.58% | 158.06 | 187.29 | 248.55 | 343.19 | 476.69 | 643.92 | 778.46 |
| | 250 | 0.00271 | 388.01 | 179.55 | 54.74% | 174.75 | 203.09 | 262.05 | 351.18 | 471.23 | 618.31 | 724.36 |
| | 300 | 0.00269 | 387.57 | 163.14 | 54.02% | 188.64 | 216.10 | 272.84 | 356.31 | 466.81 | 596.82 | 696.54 |
| | 400 | 0.00265 | 388.78 | 138.96 | 51.68% | 211.41 | 236.63 | 289.31 | 365.22 | 460.07 | 571.35 | 651.03 |
| | 500 | 0.00264 | 388.38 | 123.79 | 50.37% | 223.45 | 249.45 | 299.76 | 369.29 | 455.55 | 550.83 | 619.16 |
| **(12, 100)** | 30 | 0.00381 | 388.68 | 1093.61 | 73.51% | 32.00 | 45.15 | 84.73 | 177.26 | 391.53 | 832.54 | 1331.17 |
| | 50 | 0.00331 | 387.67 | 590.41 | 68.52% | 56.42 | 76.22 | 128.22 | 235.73 | 445.58 | 816.50 | 1185.24 |
| | 100 | 0.00291 | 387.73 | 310.00 | 61.66% | 100.04 | 126.71 | 190.41 | 302.08 | 482.64 | 745.49 | 965.87 |
| | 150 | 0.00279 | 388.71 | 246.00 | 58.60% | 133.39 | 161.14 | 224.78 | 327.45 | 481.86 | 685.43 | 854.06 |
| | 200 | 0.00274 | 387.52 | 205.98 | 56.54% | 153.72 | 183.13 | 245.16 | 341.23 | 477.15 | 644.86 | 774.87 |
| | 250 | 0.00270 | 387.63 | 181.19 | 55.08% | 171.82 | 200.99 | 260.82 | 350.06 | 473.20 | 620.73 | 730.66 |
| | 300 | 0.00268 | 387.90 | 164.55 | 54.00% | 183.94 | 212.72 | 272.09 | 354.83 | 470.10 | 601.04 | 700.70 |
| | 400 | 0.00265 | 388.06 | 140.60 | 51.80% | 207.26 | 233.93 | 287.39 | 364.67 | 462.68 | 571.49 | 651.12 |
| | 500 | 0.00265 | 388.11 | 127.13 | 50.89% | 220.56 | 247.14 | 298.20 | 367.75 | 455.02 | 553.74 | 627.42 |

A second adjustment, is to use a FAR $\alpha'' \neq \alpha$ so as the $\frac{\#\{CARL_0 < ARL_0\}}{N} < p$, i.e., the percentage of charts with $CARL_0 < ARL_0$ does not exceed $100p\%$. We will refer to this method as adjustment B (or adj-B) and has been also applied by Sarmiento et al. (2022) while it is based on the idea of exceedance probability criterion, originally suggested by Albers and Kallenberg (2004). Again, for the determination of $\alpha''$ we apply Algorithm 1 for a wide range of candidate $\alpha''$ values.

Similar to Table 3, in Table 4 we provide the adjusted $\alpha''$ values (see the respective column) along with the $AARL$, the $SDARL$, the empirical percentage of charts with $CARL_0 < 370.4$ (column '$perc$') and percentiles of the distribution of $CARL_0$. The results show that a substantial decrease in FAR is necessary (i.e., $\alpha'' < \alpha$) in order to secure that only a small percentage of charts will have an $CARL_0 < 370.4$. This constraint has an opposite effect on the distribution of $CARL_0$, which now has an increased variability; see for example the $SDARL$ values as well as the range between the 0.05- and 0.95-percentile points. This increase in variability is obvious when $m < 500$. Again, this is expected since for not very large Phase I samples, the variability of the estimates requires a generous decrease in FAR ($a''$ has to be much smaller than $\alpha$) as a protection against small $CARL_0$ values.

Also, following Sarmiento et al. (2022), we consider a slightly relaxed constraint and provide the $a''$ so as $\frac{\#\{CARL_0 < (1+\epsilon)^{-1} ARL_0\}}{N} < p$, where $\epsilon \in (0,1)$ is a tolerance parameter. Specifically, in Table 5 we provide the $a''$ for $p = 0.10$ and $\epsilon = 0.20$. That is, the percentage of charts with IC $ARL$ lower than $(1 + 0.20)^{-1} \cdot 370.4 = 308.67$ does not exceed 10%. The results are given in Table 5 and are similar to that in Table 4. Now the $a''$ values are not that small as for the case $p = 0.05$ and $\epsilon = 0$, but the $SDARL$ values remain quite large in all the considered cases. Again, as the sample size $m$ increases, the $a''$ increases as well. As expected, the percentage of charts with IC $ARL$ not larger than 308 is approximately 10%. See also that the 0.10-percentile point of the conditional IC $ARL$ distribution is very close to this value.

(Please Insert Tables 4 and 5 around Here)

**Table 4:** Adjusted FAR values (method adj-B, $\epsilon = 0$, $p = 0.05$)

| $(\theta_{01}, \theta_{02})$ | $m$ | $\alpha''$ | AARL | SDARL | perc | 5% | 10% | 25% | 50% | 75% | 90% | 95% |
|---|---|---|---|---|---|---|---|---|---|---|---|---|
| (2, 30) | 30 | 0.00003 | 245917.40 | 4182022 | 4.59% | 401.48 | 805.65 | 3091.24 | 15242.59 | 75231.77 | 312161.11 | 708898.68 |
| | 50 | 0.00017 | 12315.46 | 33560.86 | 4.99% | 371.25 | 604.32 | 1452.75 | 3965.48 | 11035.36 | 27553.06 | 47380.03 |
| | 100 | 0.00052 | 2439.11 | 2912.38 | 4.99% | 371.46 | 506.01 | 871.43 | 1598.69 | 2957.23 | 5198.91 | 7189.17 |
| | 150 | 0.00078 | 1490.04 | 1216.78 | 5.00%. | 370.41 | 472.45 | 718.68 | 1148.72 | 1866.66 | 2854.09 | 3721.17 |
| | 200 | 0.00094 | 1182.31 | 792.54 | 4.88% | 372.62 | 465.07 | 661.90 | 983.72 | 1470.64 | 2118.84 | 2633.96 |
| | 250 | 0.00109 | 988.39 | 544.74 | 4.90% | 372.70 | 448.87 | 610.74 | 859.55 | 1223.79 | 1682.12 | 2035.26 |
| | 300 | 0.00120 | 880.36 | 433.35 | 4.96% | 371.06 | 437.53 | 575.81 | 784.35 | 1083.99 | 1441.16 | 1707.91 |
| | 400 | 0.00137 | 762.50 | 313.70 | 4.97% | 370.74 | 427.94 | 540.35 | 704.36 | 918.87 | 1168.73 | 1351.10 |
| | 500 | 0.00149 | 695.98 | 251.35 | 4.99% | 370.45 | 421.57 | 517.28 | 655.01 | 826.18 | 1023.04 | 1161.59 |
| (3, 12) | 30 | 0.00004 | 167262.60 | 1636568 | 4.85% | 380.76 | 758.92 | 2648.29 | 11790.55 | 53684.27 | 215529.69 | 487976.45 |
| | 50 | 0.00018 | 11882.98 | 37002.81 | 4.78% | 378.19 | 611.16 | 1420.58 | 3727.74 | 9848.75 | 24918.89 | 43439.36 |
| | 100 | 0.00055 | 2340.42 | 2758.70 | 4.94% | 371.81 | 500.51 | 841.68 | 1534.71 | 2800.06 | 4902.30 | 6870.18 |
| | 150 | 0.00079 | 1463.79 | 1170.61 | 4.90% | 373.09 | 476.34 | 714.79 | 1141.87 | 1828.16 | 2787.58 | 3613.60 |
| | 200 | 0.00097 | 1135.31 | 729.43 | 4.93% | 371.37 | 455.54 | 643.30 | 952.70 | 1408.19 | 2019.37 | 2529.81 |
| | 250 | 0.00111 | 966.25 | 532.10 | 4.98% | 370.95 | 439.02 | 602.99 | 844.67 | 1194.73 | 1621.48 | 1978.13 |
| | 300 | 0.00121 | 874.96 | 425.72 | 4.95% | 371.67 | 440.25 | 576.69 | 786.64 | 1070.86 | 1420.59 | 1680.18 |
| | 400 | 0.00138 | 755.30 | 308.39 | 4.83% | 372.67 | 427.75 | 536.77 | 696.57 | 910.15 | 1155.81 | 1336.91 |
| | 500 | 0.00150 | 686.43 | 245.24 | 4.98% | 370.79 | 419.78 | 513.24 | 644.83 | 811.98 | 1010.69 | 1140.62 |
| (12, 100) | 30 | 0.00002 | 362720.50 | 2828338 | 4.32% | 429.47 | 938.91 | 3964.36 | 22293.85 | 118023.69 | 502707.40 | 1214138.79 |
| | 50 | 0.00015 | 14341.42 | 40746.72 | 4.88% | 375.99 | 625.85 | 1561.49 | 4465.47 | 12621.96 | 31799.31 | 55838.01 |
| | 100 | 0.00049 | 2626.79 | 2995.07 | 4.92% | 372.91 | 518.81 | 907.90 | 1721.10 | 3231.65 | 5564.25 | 7853.49 |
| | 150 | 0.00076 | 1512.52 | 1209.07 | 4.90% | 372.94 | 476.12 | 729.93 | 1176.38 | 1896.12 | 2938.41 | 3769.61 |
| | 200 | 0.00094 | 1183.51 | 785.58 | 4.99% | 370.76 | 458.75 | 655.49 | 988.31 | 1485.28 | 2128.41 | 2642.65 |
| | 250 | 0.00108 | 1004.97 | 560.83 | 4.86% | 372.91 | 451.28 | 616.16 | 876.37 | 1250.31 | 1714.68 | 2045.14 |
| | 300 | 0.00119 | 895.96 | 445.43 | 4.95% | 371.22 | 439.87 | 585.15 | 803.32 | 1101.69 | 1462.34 | 1738.95 |
| | 400 | 0.00135 | 774.23 | 323.62 | 4.95% | 371.11 | 429.53 | 547.66 | 711.61 | 934.08 | 1190.38 | 1379.25 |
| | 500 | 0.00147 | 704.32 | 256.05 | 4.88% | 372.21 | 423.38 | 521.32 | 660.21 | 838.32 | 1041.39 | 1188.36 |

**Table 5:** Adjusted FAR values (method adj-B, $\epsilon = 0.20$, $p = 0.10$)

| $(\theta_{01}, \theta_{02})$ | $m$ | $\alpha''$ | AARL | SDARL | perc | 5% | 10% | 25% | 50% | 75% | 90% | 95% |
|---|---|---|---|---|---|---|---|---|---|---|---|---|
| **(2, 30)** | 30 | 0.00016 | 23739.12 | 164360.50 | 10.00% | 171.43 | 308.70 | 918.30 | 3309.14 | 12113.23 | 39092.81 | 81517.50 |
| | 50 | 0.00046 | 3857.84 | 9538.12 | 9.92% | 202.44 | 310.98 | 655.74 | 1547.81 | 3804.84 | 8501.12 | 13893.78 |
| | 100 | 0.00097 | 1250.54 | 1270.07 | 9.89% | 234.58 | 309.87 | 503.36 | 876.34 | 1536.78 | 2570.10 | 3486.27 |
| | 150 | 0.00128 | 877.09 | 653.03 | 9.90% | 248.91 | 389.95 | 458.16 | 704.47 | 1088.26 | 1623.06 | 2099.49 |
| | 200 | 0.00150 | 725.49 | 437.07 | 9.89% | 254.62 | 308.74 | 427.77 | 620.60 | 899.52 | 1268.72 | 1560.89 |
| | 250 | 0.00167 | 639.09 | 326.79 | 9.94% | 261.11 | 309.29 | 411.10 | 567.86 | 783.11 | 1053.59 | 1259.58 |
| | 300 | 0.00178 | 593.31 | 273.91 | 9.88% | 265.95 | 309.27 | 400.06 | 538.17 | 724.37 | 944.86 | 1110.95 |
| | 400 | 0.00196 | 528.31 | 203.48 | 9.99% | 269.87 | 308.74 | 382.26 | 492.18 | 631.61 | 793.07 | 910.83 |
| | 500 | 0.00207 | 494.85 | 165.70 | 9.86% | 275.21 | 309.47 | 376.10 | 471.52 | 584.43 | 713.34 | 803.00 |
| **(3, 12)** | 30 | 0.00017 | 24679.48 | 221580.20 | 9.94% | 172.55 | 311.03 | 893.24 | 3062.28 | 10981.93 | 36682.46 | 76706.78 |
| | 50 | 0.00046 | 3766.15 | 8835.48 | 9.78% | 204.71 | 313.15 | 657.58 | 1545.42 | 3709.46 | 8271.28 | 13373.97 |
| | 100 | 0.00101 | 1211.14 | 1440.02 | 9.99% | 236.18 | 308.93 | 495.01 | 844.84 | 1468.13 | 2423.28 | 3313.65 |
| | 150 | 0.00132 | 852.35 | 617.26 | 9.84% | 248.69 | 310.43 | 450.59 | 684.27 | 1060.09 | 1588.64 | 2008.55 |
| | 200 | 0.00152 | 715.06 | 428.13 | 9.96% | 255.07 | 309.07 | 426.52 | 610.64 | 884.45 | 1238.72 | 1517.74 |
| | 250 | 0.00167 | 632.91 | 323.00 | 9.99% | 261.31 | 308.75 | 409.47 | 562.52 | 772.47 | 1043.35 | 1247.17 |
| | 300 | 0.00180 | 583.52 | 268.64 | 9.86% | 265.86 | 309.62 | 398.06 | 529.00 | 704.59 | 918.97 | 1081.39 |
| | 400 | 0.00196 | 528.42 | 201.68 | 9.85% | 272.56 | 309.62 | 385.85 | 491.42 | 634.60 | 789.05 | 900.40 |
| | 500 | 0.00211 | 488.71 | 162.54 | 9.91% | 275.80 | 309.10 | 372.86 | 462.00 | 575.64 | 700.59 | 789.32 |
| **(12, 100)** | 30 | 0.00013 | 28016.28 | 193847.90 | 9.64% | 178.19 | 320.89 | 987.75 | 3815.83 | 14730.15 | 50409.46 | 102190.44 |
| | 50 | 0.00043 | 3989.15 | 8332.86 | 9.90% | 200.11 | 310.56 | 666.32 | 1636.65 | 4020.54 | 9061.02 | 14648.01 |
| | 100 | 0.00097 | 1264.93 | 1273.59 | 9.93% | 230.33 | 309.61 | 504.94 | 881.11 | 1567.44 | 2608.93 | 3506.85 |
| | 150 | 0.00127 | 893.98 | 665.70 | 9.86% | 245.07 | 310.18 | 458.59 | 713.90 | 1118.45 | 1679.25 | 2129.28 |
| | 200 | 0.00147 | 743.06 | 455.64 | 9.81% | 254.82 | 310.39 | 434.65 | 634.11 | 921.57 | 1292.91 | 1619.24 |
| | 250 | 0.00166 | 640.87 | 330.25 | 9.96% | 258.85 | 309.08 | 410.11 | 567.55 | 790.98 | 1060.20 | 1266.49 |
| | 300 | 0.00178 | 589.52 | 272.28 | 9.98% | 264.37 | 308.79 | 398.24 | 533.78 | 716.09 | 933.66 | 1103.75 |
| | 400 | 0.00194 | 535.70 | 209.45 | 9.90% | 270.95 | 309.21 | 387.27 | 497.60 | 640.66 | 808.14 | 929.88 |
| | 500 | 0.00206 | 500.37 | 171.23 | 9.80% | 275.77 | 309.57 | 377.18 | 473.13 | 590.80 | 724.01 | 819.80 |

## 3.3 Out-of-Control Performance of the SH$_K$-chart in Case K and Case U

Next, we investigate the OOC performance of the SH$_K$-chart for separate (i.e., not simultaneous) shifts in $\theta_{01}$ and $\theta_{02}$, under Case K and Case U. We believe that since we are using only one control chart then we should examine the OOC performance of the SHK-chart by considering separate shifts in process parameters and the case of simultaneous shifts could be treated at least with the use of two separate charts, one of each parameter.

We consider the cases $m = 100$ and 200, as well as different pairs of limits and calculate, for a range of shifts in process parameters, the $AARL$. The OOC values of the process parameters are defined as $\theta_{11} = \delta_1 \theta_{01}$ and $\theta_{12} = \delta_2 \theta_{02}$ where $\delta_1, \delta_2 > 0$. The considered shifts are in $\{0.5, 0.6, \ldots, 0.9, 1, 1.1, 1.2, \ldots, 2.0\}$ whereas for $\delta_1 = \delta_2 = 1$, the process is IC.

In Figures 2-4 and 5-7 we provide the OOC $AARL$ values for each shift in $\theta_{01}$ and $\theta_{02}$, respectively, for different pairs of control limits, which are obtained under the different adjustment methods. Note also that in Case K, we simply calculate the OOC $ARL$, say $ARL_1$, as:

$$ARL_1 = \frac{1}{1 - F_K(UCL; \theta_{11}, \theta_{12}) + F_K(LCL; \theta_{11}, \theta_{12})}, \qquad (12)$$

Recall that both Lima-Filho et al. (2020) and Lima-Filho and Bayer (2021) focused on additive changes only in the process mean level. However, a change in exactly one of the process parameters, affects both the process mean level and the variability of the process.

Except for the OOC $ARL$ in Case K, in Figures 2-7 we provide also the $AARL$ values in Case U, for four different pairs of estimated control limits: The plug-in limits, the limits obtained using Adjustment method A (see 'Adj-A'), the limits obtained using Adjustment method B for $\epsilon = 0$ and $p = 0.05$ (see 'Adj-B, eps=0') and the limits obtained using Adjustment method B for $\epsilon = 0.20$ and $p = 0.10$ (see 'Adj-B, eps=0.20').

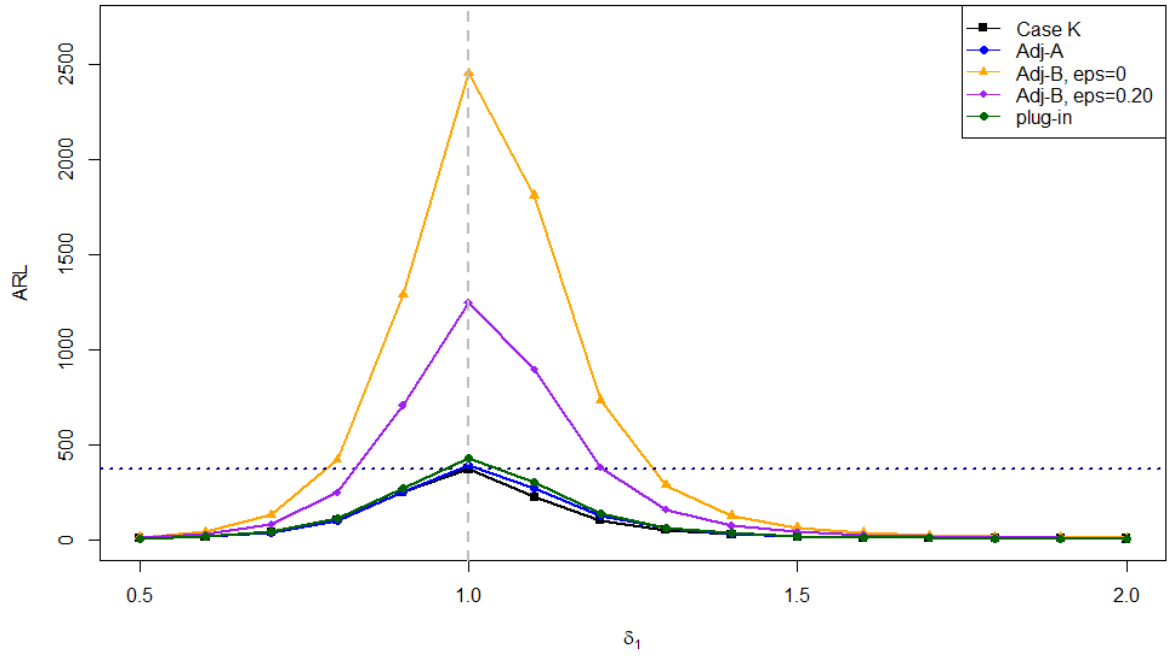

(a) $m = 100$

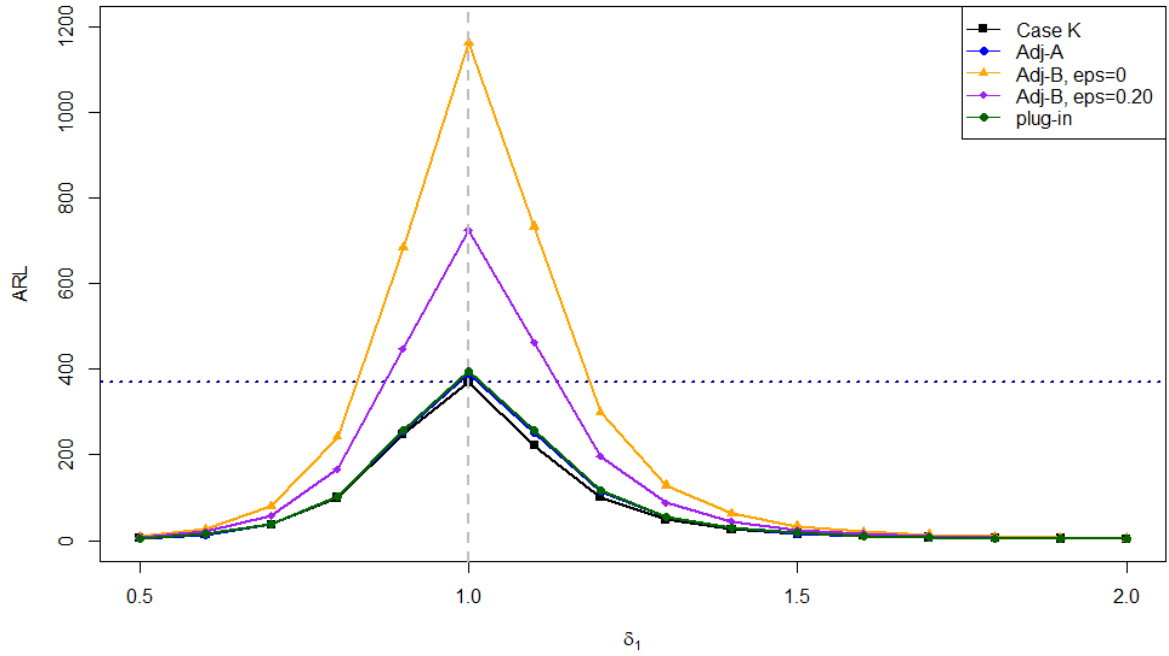

(b) $m = 200$

**Figure 2:** OOC performance of the $SH_K$-chart (changes in $\theta_{01}$) in Case K and in Case U, for four different pairs of estimated control limits, $\alpha = 0.0027$, $(\theta_{01}, \theta_{02}) = (2, 30)$.

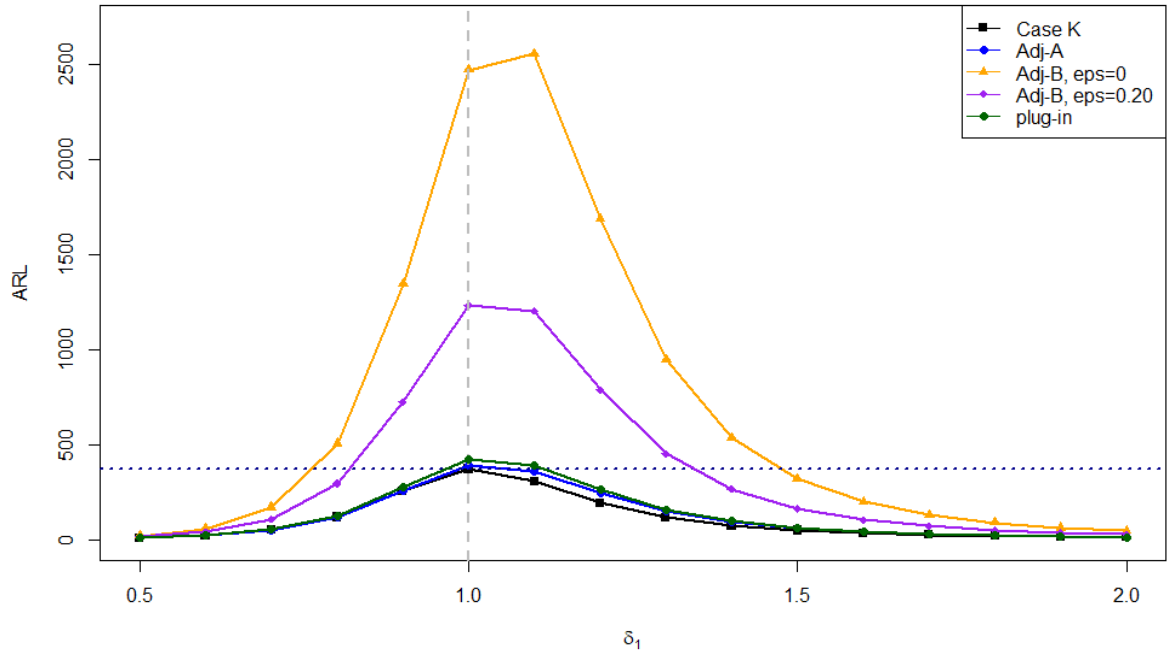

(a) $m = 100$

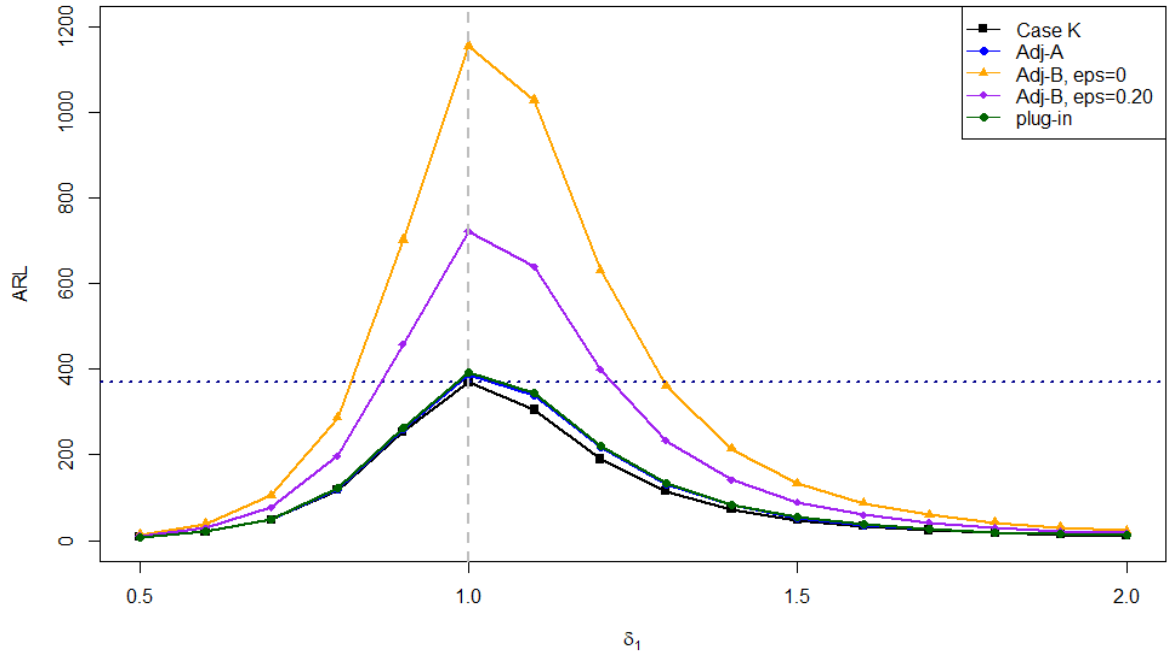

(b) $m = 200$

**Figure 3:** OOC performance of the $SH_K$-chart (changes in $\theta_{01}$) in Case K and in Case U, for four different pairs of estimated control limits, $\alpha = 0.0027$, $(\theta_{01}, \theta_{02}) = (3,12)$.

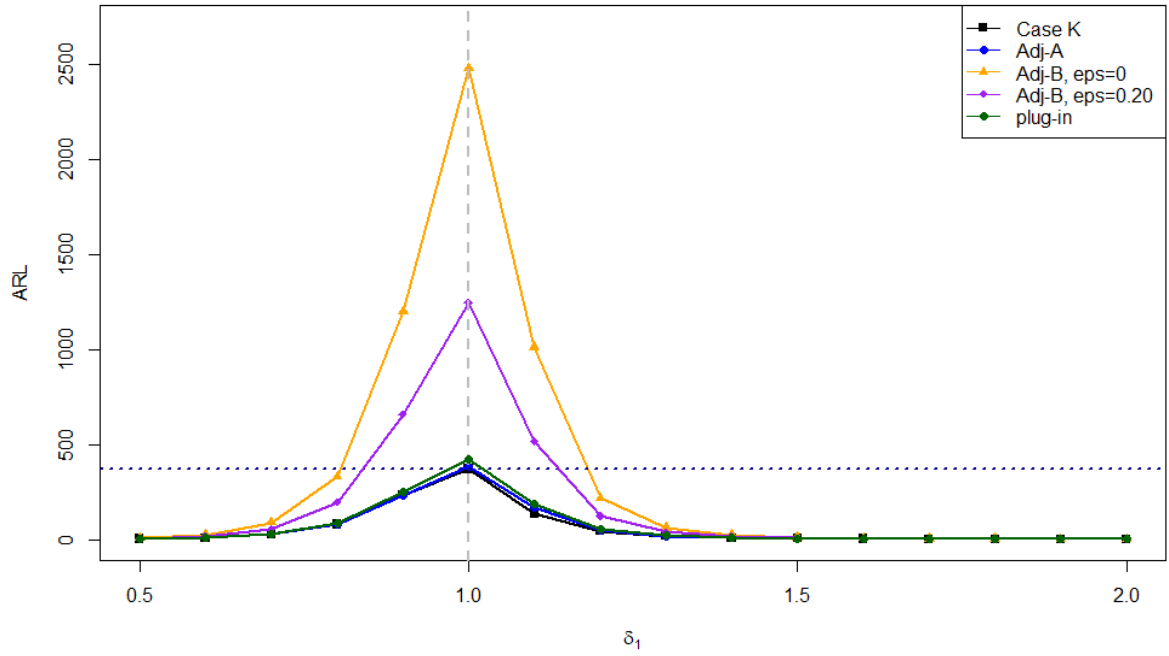

(a) $m = 100$

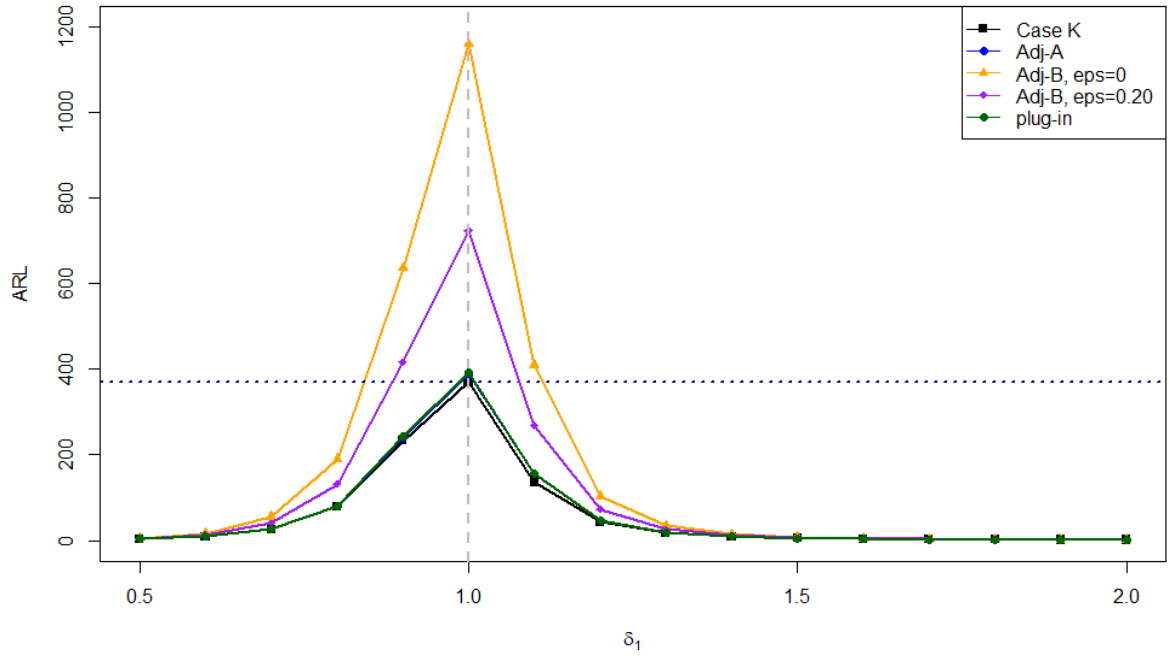

(b) $m = 200$

**Figure 4:** OOC performance of the $SH_K$-chart (changes in $\theta_{01}$) in Case K and in Case U, for four different pairs of estimated control limits, $\alpha = 0.0027$, $(\theta_{01}, \theta_{02}) = (12, 100)$.

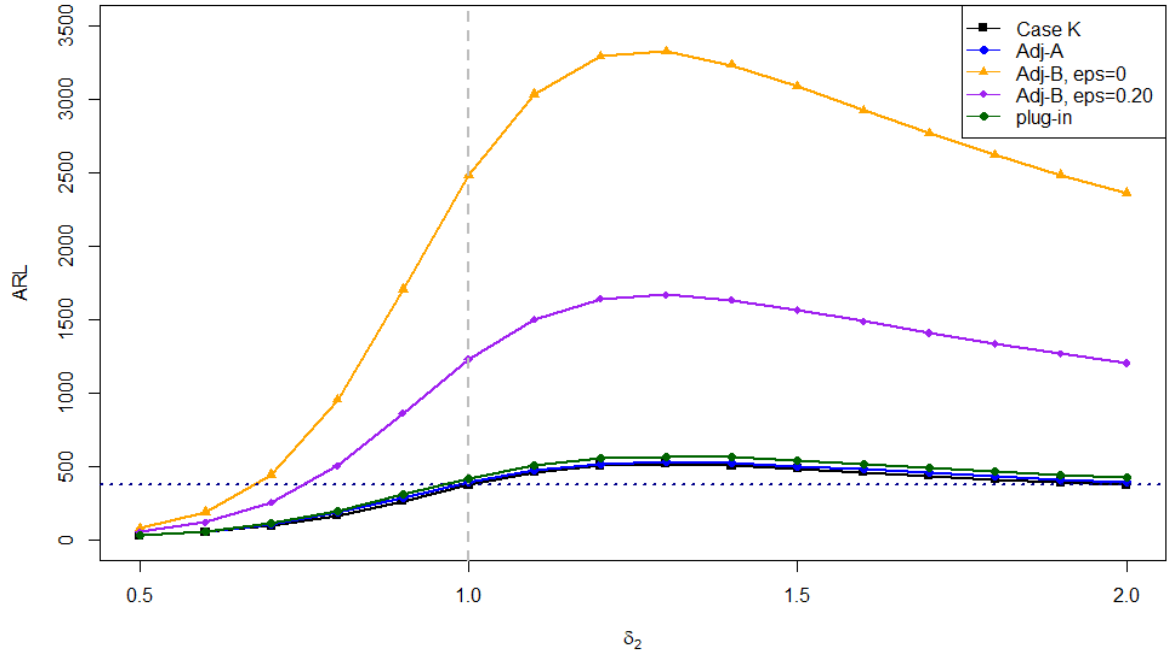

(a) $m = 100$

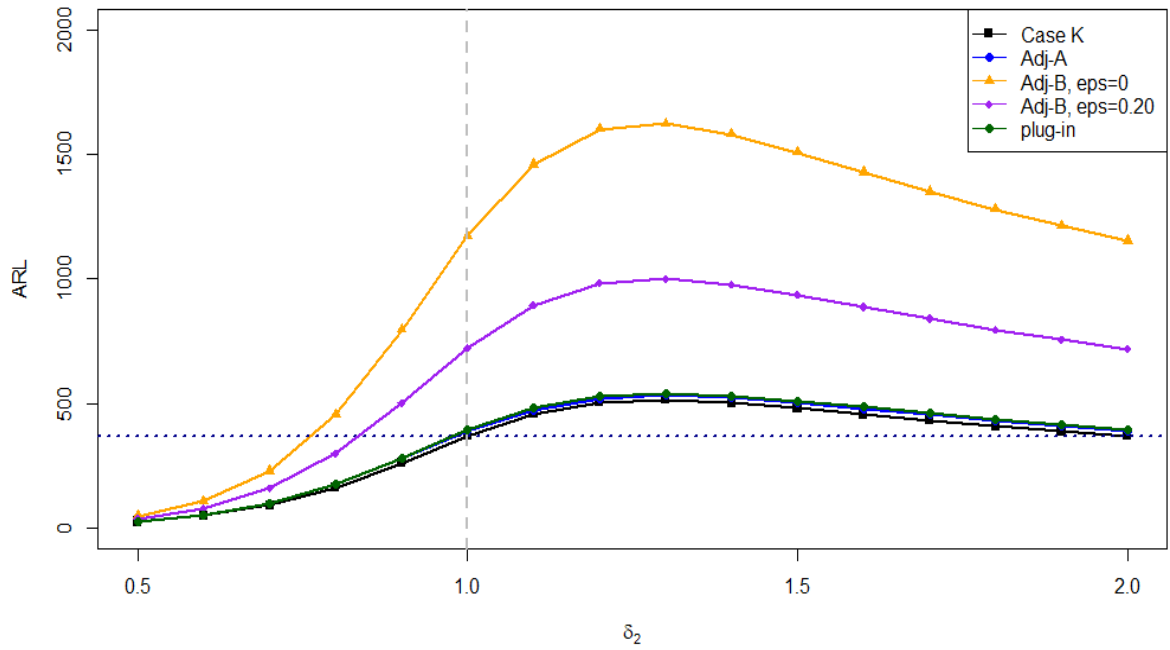

(b) $m = 200$

**Figure 5:** OOC performance of the $SH_K$-chart (changes in $\theta_{02}$) in Case K and in Case U, for four different pairs of estimated control limits, $\alpha = 0.0027$, $(\theta_{01}, \theta_{02}) = (2, 30)$.

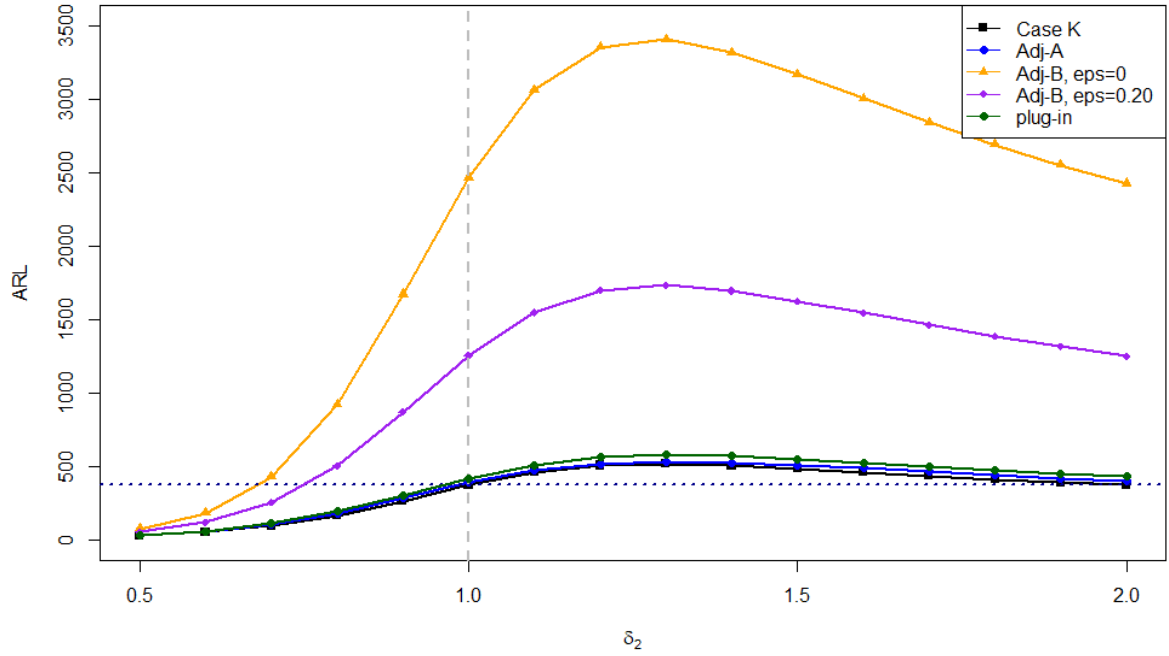

(a) $m = 100$

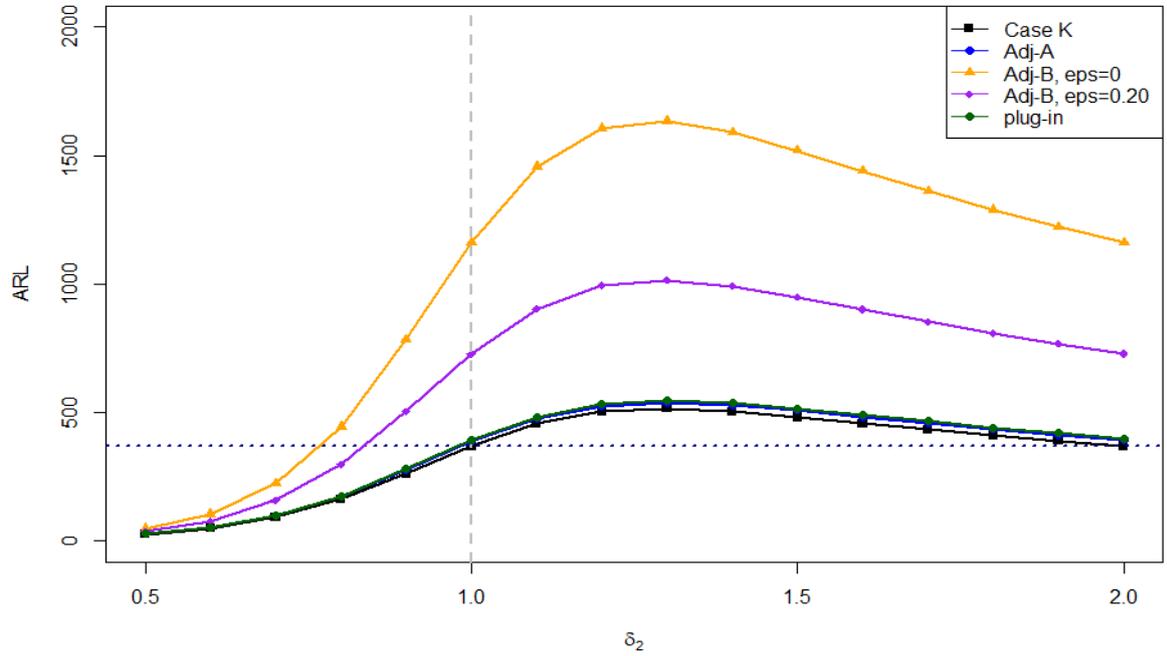

(b) $m = 200$

**Figure 6:** OOC performance of the $SH_K$-chart (changes in $\theta_{02}$) in Case K and in Case U, for four different pairs of estimated control limits, $\alpha = 0.0027$, $(\theta_{01}, \theta_{02}) = (3, 12)$, $m = 100$ (left panel) and $m = 200$ (right panel).

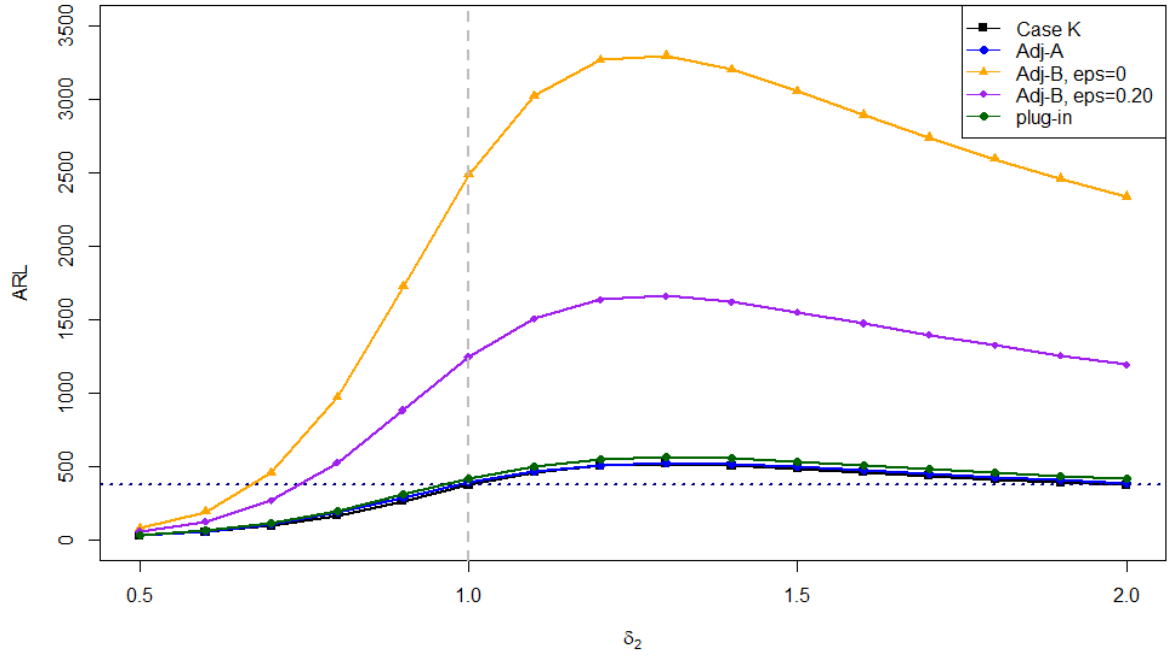

(a) $m = 100$

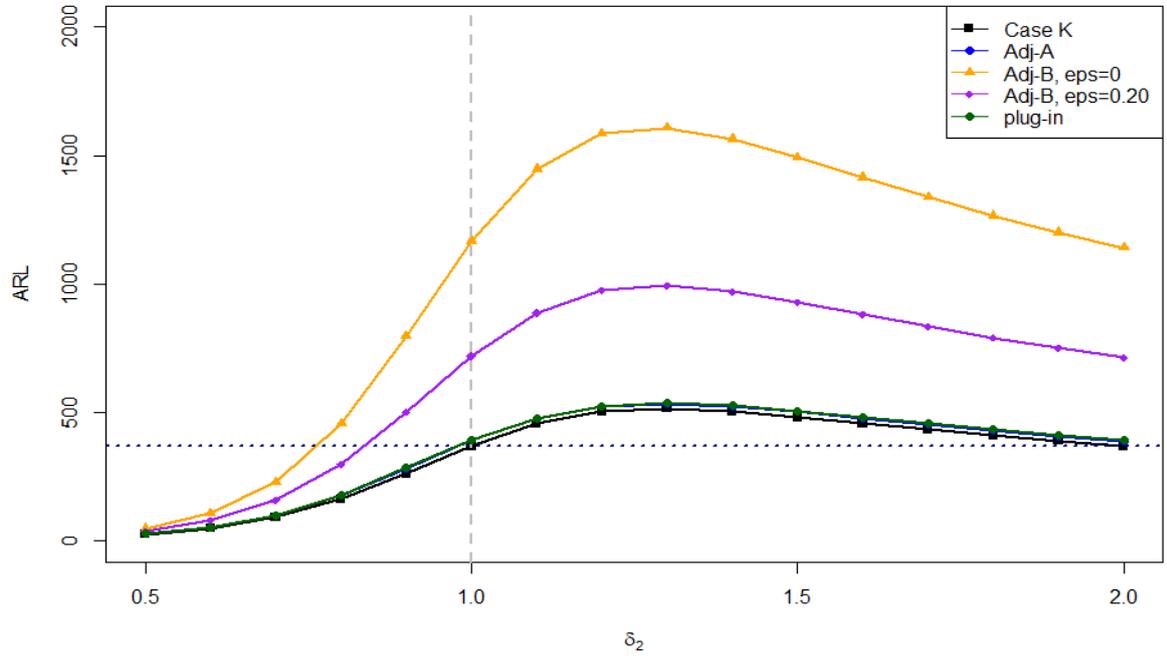

(b) $m = 200$

**Figure 7:** OOC performance of the SH$_K$-chart (changes in $\theta_{02}$) in Case K and in Case U, for four different pairs of estimated control limits, $\alpha = 0.0027$, $(\theta_{01}, \theta_{02}) = (12, 100)$, $m = 100$ (left panel) and $m = 200$ (right panel).

The results show that when the limits are determined under the adjustment B, then there is an effect on the OOC $AARL$ values since they are much larger than the other two methods, especially for shifts $\delta_1 \in [0.8,1.4]$. Clearly, this adjustment gives protection against designs with increased FAR but reduces the power of the chart in the detection of small to moderate increasing shifts in $\theta_{01}$. When $m$ increases, these differences become smaller but as already stated, the larger the Phase I sample, the better (and more accurate) estimates for process parameters.

Another interesting finding, that applies in both Case U and Case K is that the considered SH$_K$-chart cannot detect increasing shifts, at least for $\delta_2 \in (1,2]$. Probably this is attributed to the use of equal tail probability limits and an $ARL$-unbiased design could be a solution. However, this is left as a topic for future research.

Before closing this section, we provide in Figures 8-10 the boxplots of the OOC $ARL$ values, given the estimates $\widehat{LCL}$ and $\widehat{UCL}$ of the control limits, under the four methods to obtain the estimated limits and for specific shifts in process parameters. For illustrative purposes we provide the respective figures for $\alpha = 0.0027$, $(\theta_{01}, \theta_{02}) = (3,12)$, $m = 200$ and $(\delta_1, \delta_2) \in \{(0.8,1), (1.2,1), (1,0.8)\}$. The red vertical line denotes the $ARL_1$ value in Case K and as we easily notice, when the Adj-A or the plug-limits are used, the median of the distribution is close to this (theoretical) value. However, for both cases there is clearly a positive skewness which means that several practitioners will end with much larger OOC $ARL$ values than the theoretical one. On the other hand, under the Adj-B design method we notice the vast majority of the practitioners will end up with much larger values than the theoretical one, which is due to the wider control limits that are used. Clearly, there is no optimal solution and practitioners are advised to determine the best design that fits their needs.

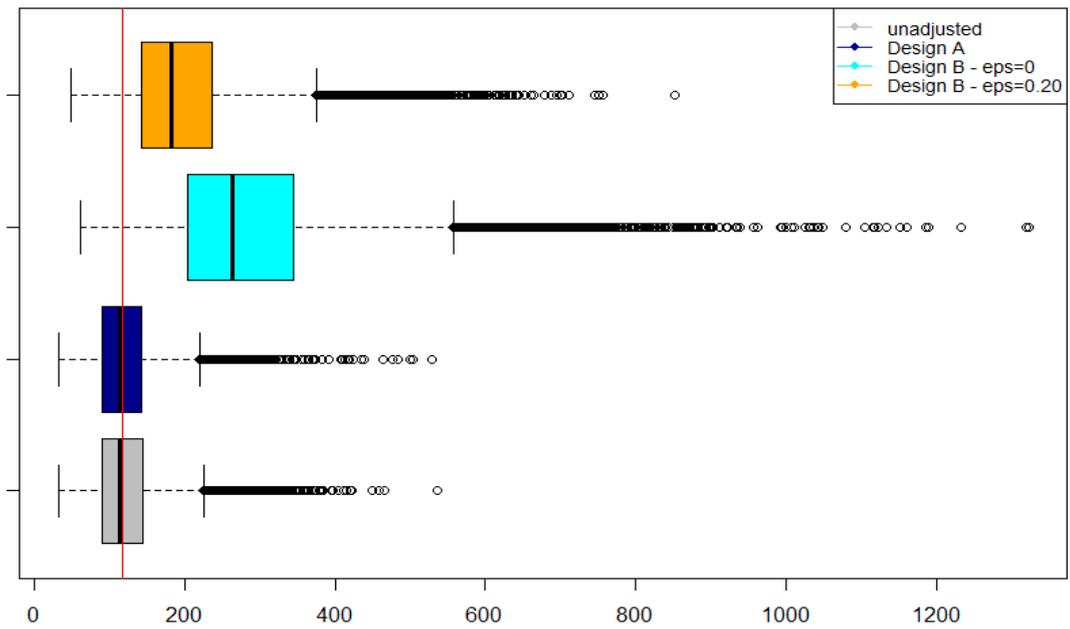

**Figure 8:** Boxplots with the OOC conditional ARL values, for $\alpha = 0.0027$, $(\theta_{01}, \theta_{02}) = (3,12)$, $m = 200$ and $(\delta_1, \delta_2) = (0.8,1)$.

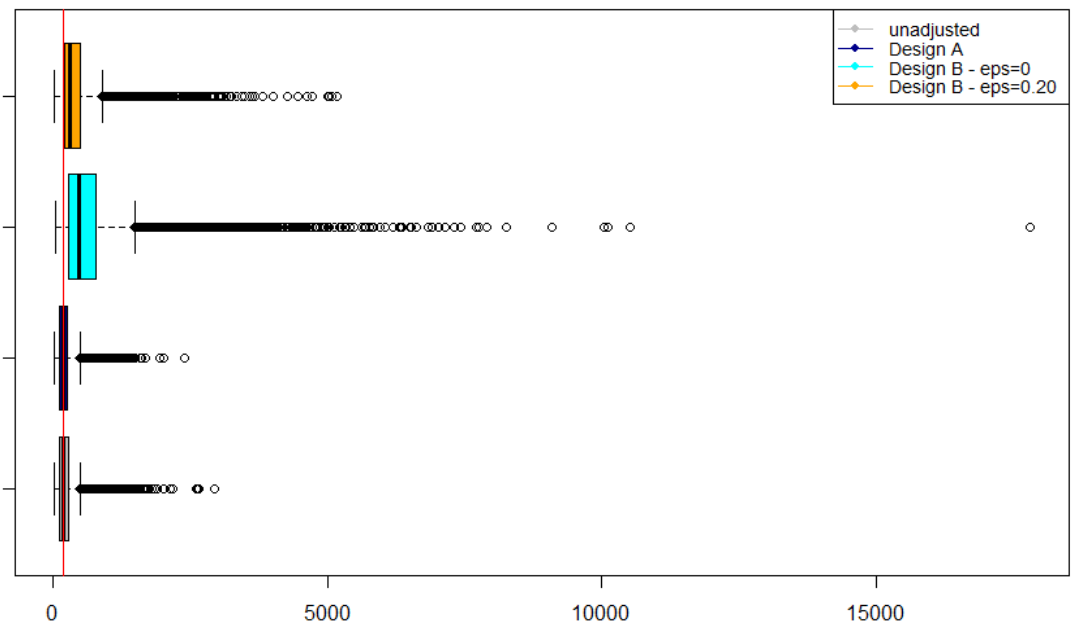

**Figure 9:** Boxplots with the OOC conditional ARL values, for $\alpha = 0.0027$, $(\theta_{01}, \theta_{02}) = (3,12)$, $m = 200$ and $(\delta_1, \delta_2) = (1.2,1)$.

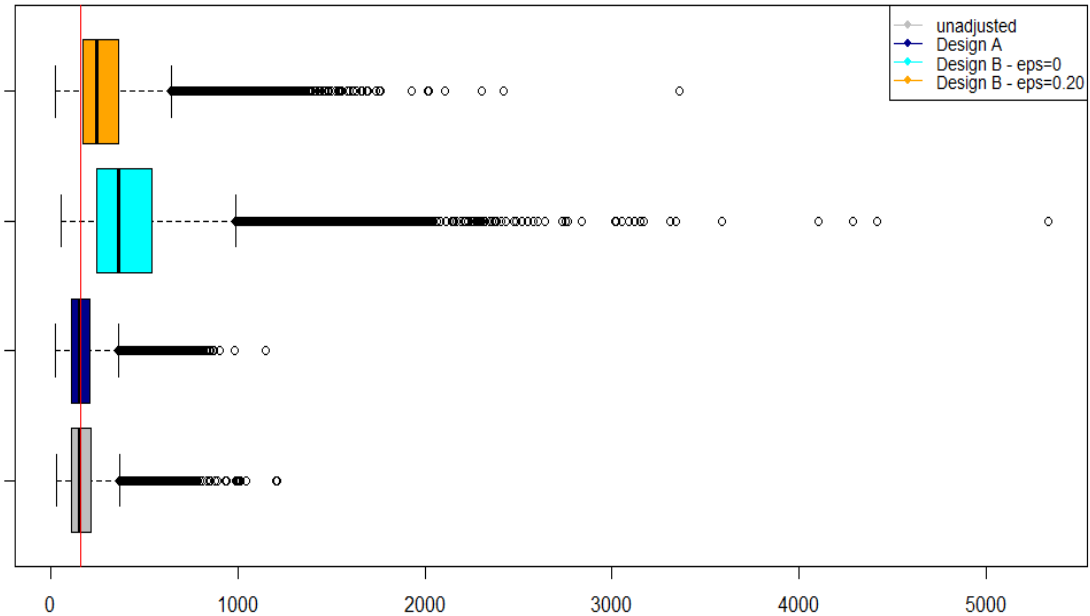

**Figure 10:** Boxplots with the OOC conditional ARL values, for $\alpha = 0.0027$, $(\theta_{01}, \theta_{02}) = (3,12)$, $m = 200$ and $(\delta_1, \delta_2) = (1, 0.8)$.

## 4. Applications

In this section, we present two illustrative examples regarding the usefulness of the Kumaraswamy distribution in modelling and monitoring data from the open unit interval (0, 1) as well as the how to determine the control limits of a two-sided chart when the process parameters are estimated.

### 5.1 Monitoring the Proportions of Unconverted Mass – Simulated data

Let us consider the case of a radial tire manufacturing process of a multinational company of rubber products. The interest is on monitoring the proportion of unconverted mass which is the rate between the volume of raw material that was not converted into product and the total volume, being directly related to the loss of raw material. At the $i^{th}$ inspection this proportion is recorded, and we assume this number comes from a Kumaraswamy distribution. This setting resembles the one in Section 4.1 of Bayer et al. (2018). For monitoring the process, we need first to estimate the process parameters and then setup the two-sided $SH_K$-chart.

The numerical study in 3.1 showed that large Phase I samples are needed. Thus, we assume that the available Phase I sample is the one given in Table 6 (the values are given by row) where $m = 100$. The values have been simulated from the $Kuma(2,350)$ distribution, which is the IC model. However, the practitioner does not know the true $\theta_{01}, \theta_{02}$ values.

**Table 6:** Simulated data from $Kuma(2,350)$ distribution – Phase I samples

| | | | | | | | | | |
|---|---|---|---|---|---|---|---|---|---|
| 0.0257 | 0.0397 | 0.0694 | 0.0757 | 0.0994 | 0.0974 | 0.0343 | 0.0468 | 0.0307 | 0.0539 |
| 0.0071 | 0.0632 | 0.0240 | 0.0131 | 0.0444 | 0.0533 | 0.0110 | 0.0467 | 0.0450 | 0.0174 |
| 0.0868 | 0.0509 | 0.0351 | 0.0456 | 0.0554 | 0.0082 | 0.0202 | 0.0321 | 0.0376 | 0.0478 |
| 0.0190 | 0.0197 | 0.0910 | 0.0264 | 0.0526 | 0.0636 | 0.0436 | 0.0849 | 0.0765 | 0.0110 |
| 0.0429 | 0.0465 | 0.0343 | 0.0484 | 0.0596 | 0.0415 | 0.0071 | 0.0354 | 0.1105 | 0.0372 |
| 0.0472 | 0.0394 | 0.0086 | 0.0384 | 0.0465 | 0.0369 | 0.0205 | 0.0680 | 0.0457 | 0.0150 |
| 0.0488 | 0.0401 | 0.0506 | 0.0142 | 0.0673 | 0.0349 | 0.0290 | 0.0212 | 0.0319 | 0.0769 |
| 0.0740 | 0.0509 | 0.0684 | 0.0341 | 0.0503 | 0.0901 | 0.0403 | 0.0064 | 0.0832 | 0.0287 |
| 0.0549 | 0.0715 | 0.0329 | 0.0236 | 0.0756 | 0.0305 | 0.0468 | 0.0071 | 0.0516 | 0.0099 |
| 0.0443 | 0.0294 | 0.0226 | 0.0523 | 0.0703 | 0.0417 | 0.0498 | 0.0334 | 0.0477 | 0.0725 |

Using these values, we estimate the (unknown) process parameters via the MLE method and the estimates of the process parameters are (in two decimals accuracy) $\hat{\theta}_1 = 2.01$ (0.16) and $\hat{\theta}_2 = 405.60$ (185.77). In parentheses we provide the standard errors.

Next, we calculate the different pairs of control limits, using the above estimates as the true parameter values whereas, we apply also the necessary adjustments on the FAR. The nominal FAR is $\alpha = 0.0027$. Therefore, the plug-in control limits, see equation (9), are equal to (in 6 decimals accuracy):

$$LCL = 0.001866, \ UCL = 0.128041.$$

Then, we calculate the control limits using the adjustments discussed in Section 3.2. First, if we want to have an $AARL$ that differs no more than 5% to the nominal value 370.4 when $m = 100$, we have to use a FAR equal to $\alpha' \approx 0.00291$; See the respective results in Table 2. Therefore, the limits are:

$$LCL_1 = 0.001937, \ UCL_1 = 0.127322.$$

If we want to have at most 5% of charts with IC performance below the nominal IC $ARL$, which is 370.4, the results in Table 4 suggest to use a FAR from 0.00049 to 0.00055. Thus, we use $\alpha'' = 0.00052$ and the respective control limits are:

$$LCL_2 = 0.000821, \quad UCL_2 = 0.142913.$$

Also, if we want only a 10% of the charts to have an IC $ARL$ below the $(1 + \dot{0.20})^{-1} \cdot 370.4 = 308.67$, we use a FAR approximately equal to 0.000983; See the results in Table 5, where the suggested FAR is from 0.00097 to 0.00101. The control limits in this case are

$$LCL_3 = 0.001128, \quad UCL_3 = 0.137363.$$

Clearly, the last two pairs of control limits are wider than the plug-in limits or the limits using the adjustment based on the $AARL$ criterion. However, this is necessary if we want to guarantee that only a small percentage of charts will have an increased false alarm rate.

In Figure 11, we provide the $SH_K$-chart chart for the Phase I data, along with the four pairs of control limits. The central line of the chart is $CL = 0.041786$. No point exceeds the control limits (for all the provided pairs) and thus we assume that the process during this period was IC. Practitioners can now proceed to Phase II analysis and the online monitoring of their process. They can choose the pair of limits according to their needs and by taking into account the findings in Section 3.4.

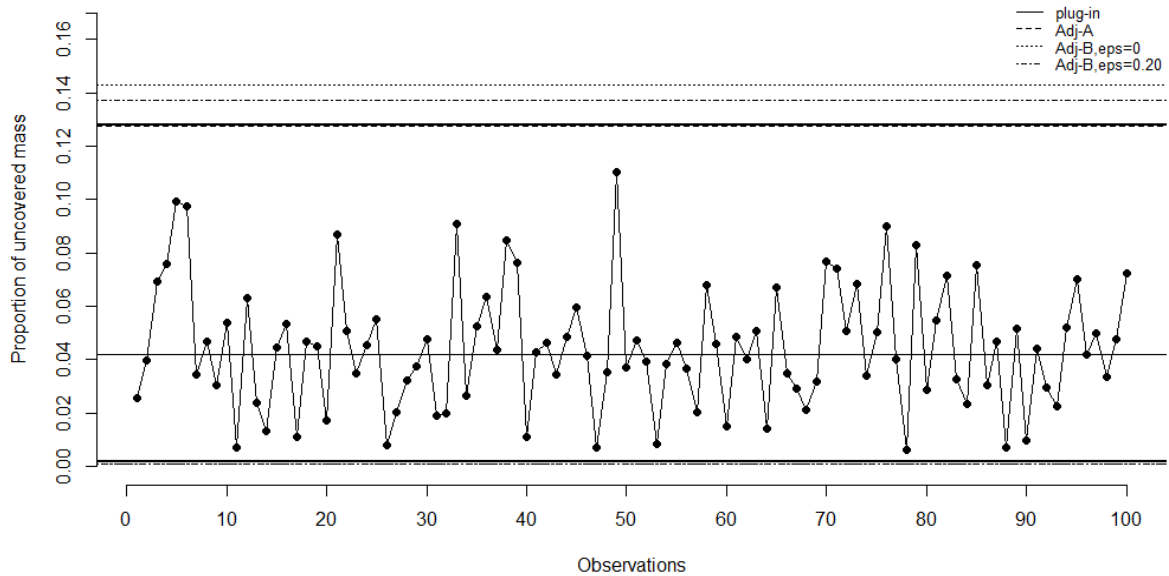

**Figure 11:** Phase I $SH_K$-chart for the simulated data in Table 5.

## 5.2 Monitoring the Relative Humidity – Real data

In this section we present a practical example using the yearly measurements of relative humidity (RH) in the Iowa of USA. These data have been also used by Lima-Filho and Bayer (2021). Here, we repeat their analysis by considering the effects of estimated parameters in the design of the chart. Specifically, we use the minimum RH from 1938 to 1967 as the Phase I data, thus $m = 30$, and the MLEs of the process parameters are $\hat{\theta}_1 = 5.631625$ (0.5967458) and $\hat{\theta}_2 = 13815.307376$ (13371.4712). In the parentheses we provide the estimated standard errors. Since we have yearly data, we use a FAR $\alpha = 0.05$ and (theoretically) we expect a false alarm signal, on average, every 20 years. Similar to the case of the previous example, we consider the same four different pairs of control limits; see Table 7. The $CL = 0.172401$ and in Figure 12 we give the $SH_k$-chart for the Phase I data.

Table 7: Different pairs of control limits

| Method | FAR | LCL | UCL |
|---|---|---|---|
| Plug-in | 0.05000 | 0.095789 | 0.231980 |
| Adj-A | 0.04803 | 0.095099 | 0.232427 |
| Adj-B, $\epsilon = 0$ | 0.00868 | 0.070062 | 0.248542 |
| Adj-B, $\epsilon = 0.20$ | 0.01854 | 0.080204 | 0.242000 |

Except for the case where the limits have been determined under the Adjustment B (with $\epsilon = 0$ and $p = 0.05$), in all other cases there is an OOC signal at point 19, which is below the lower control limit. This corresponds to the minimum relative humidity in the year 1956, which is also the minimum values in the Phase I data. Since there are no further information, we assume that this is a false alarm and proceed with the Phase II analysis with the obtained control limits.

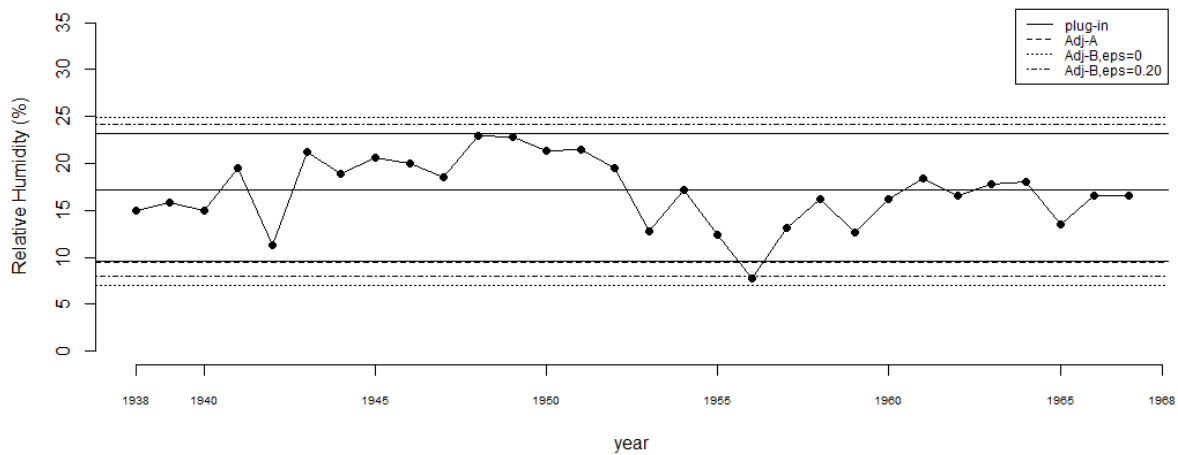

**Figure 12:** Phase I $SH_K$-chart,

In Figure 13 we give the SH$_K$-chart for the Phase II data (yearly values, from 1968 to 2016, using the same pairs of control limits. As expected, depending on the pair of control limits we use, there are differences in the OOC signals. Specifically, there is an OOC signal in the year 1972 since this point is larger than any of the considered upper control limits. According to Lima-Filho and Bayer (2021), this is attributed to the extreme impact of the El Nino event during this year. Also, in year 1993 there is an OOC signal, compared to the plug-in $\widehat{UCL}$ or the $\widehat{UCL}$ determined under the Adjustment A. However, if we use the control limits under the Adjustment B, the chart does not signal. Finally, in the year 1998, the 31st point is above the upper control limit, except for the case where the $UCL$ is obtained under the Adjustment B for $\epsilon = 0$ and $p = 0.05$.

Again, there is not an obvious suggestion on how to determine the control limits when the process parameters are estimated, since different approaches and adjustments, result in control charts with different performance. In general, practitioners are also advised to take into account the special characteristics and features of their process before selecting the nominal FAR, the size of the Phase I sample as well as how to calculate the control limits when the process parameters are estimated.

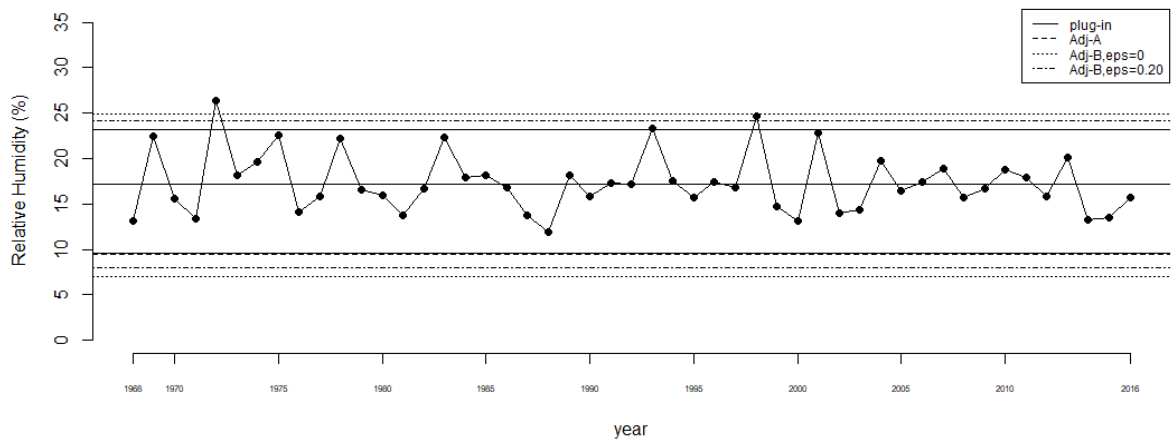

**Figure 13:** Phase II SH$_K$-chart

## 5 Conclusions

In this work, we considered the Kumaraswamy distribution as the probability model to describe data in the unit interval (0,1) and proceed with a numerical investigation of the performance of a two-sided Shewhart chart for individual observations for continuous proportions, when the process parameters are estimated. We used Monte Carlo simulation and investigated the performance of the chart under the conditional perspective, in an attempt to assist practitioners to determine the size of the Phase I sample, during the calibration stage of the control chart.

However, the numerical results showed that very large Phase I samples are necessary, i.e. with at least 2000 preliminary observations in order to have a significantly reduced variability in the conditional distribution of the IC *ARL*. In addition, even in the case of such large Phase I samples, there is a significant percentage of charts that have an increased FAR, or equivalently, a smaller IC *ARL* value than the desired one.

In an attempt to provide more reliable designs in Case U, we considered different methods for adjusting the control limits through an appropriate adjustment on the nominal FAR value, when the size of the Phase I sample is pre-specified. The results showed that Phase I samples with size lower than 100 preliminary observations are not suggested due to the increased variability of the conditional IC *ARL* distribution. Thus, practitioners have general guidelines about the size of the Phase I samples as well as whether the control limits must be adjusted or not in Case U. However, in all the considered cases the OOC performance of the chart is affected in order to guarantee its IC performance.

Topics for future research consist of the case of investigating the performance of different types of control charts for continuous proportions when rational subgroups of size $n \geq 2$, instead of individual observations, are available, either when the process are known or unknown. In addition, the case of simultaneous shifts needs to be carefully investigated, since a follow-up procedure is necessary in order to identify which of the process parameter(s) has changed.

Finally, for all computation the R programming language (R Core Team (2024)) has been used and the programs are available from the author upon request.